\documentclass[journal, 10pt]{IEEEtran}
\usepackage[reqno]{amsmath}
\usepackage{graphicx}
\usepackage{bbm}
\usepackage{mathrsfs}
\usepackage{stmaryrd}
\usepackage{graphics}
\usepackage{acronym}
\usepackage{longtable}
\usepackage{mathtools}
\usepackage{times}
\usepackage{setspace}
\usepackage{cite}
\usepackage{array}
\usepackage{subfigure}
\usepackage{amsmath,amsthm}
\usepackage{amssymb}
\usepackage{wasysym,url}
\usepackage{fixltx2e,amsmath}
\usepackage{setspace,float}
\usepackage{color}
\usepackage{cases,bm}
\usepackage{mathrsfs}
\usepackage{enumitem}
\usepackage{tikz}
\usepackage{hyperref}
\usepackage{mathtools,cuted}
\usepackage{multirow}
\usepackage{array}
\usepackage{booktabs}
\usepackage[noabbrev]{cleveref}
\crefformat{figures}{(#2#1#3)}
\usepackage{xcolor}
\usepackage[linesnumbered,ruled,vlined]{algorithm2e}
\usepackage{pdfpages}
\DontPrintSemicolon

\makeatletter
\newcommand{\leqnomode}{\tagsleft@true}
\newcommand{\reqnomode}{\tagsleft@false}
\makeatother

\newcommand{\by}{\mathbf{y}}
\newcommand{\bh}{\mathbf{h}}
\newcommand{\bx}{\mathbf{x}}
\newcommand{\bu}{\mathbf{u}}

\newcommand{\bF}{\mathbf{F}}

\newcommand{\bG}{\mathbf{G}}

\newcommand{\rr}{\mathbb{R}}
\newcommand{\zz}{\mathbb{Z}}
\newcommand{\nn}{\mathbb{N}}
\newcommand{\cc}{\mathbb{C}}

\newcommand{\TT}{\mathsf{T}}
\newcommand{\HH}{\mathsf{H}}
\newcommand{\dd}{\mathrm{d}}
\newcommand{\jj}{\mathrm{j}}

\newcommand{\sT}{\mathscr{T}}
\newcommand{\sE}{\mathscr{E}}

\newcommand{\bbeta}{\boldsymbol{\beta}}

\newcommand{\bPsi}{\boldsymbol{\Psi}}
\newcommand{\zerovec}{\boldsymbol{0}}

\newcommand{\rank}{\mathrm{rank}}

\newcommand{\Ex}{\mathbb{E}}
\newcommand{\varn}{\mathrm{var}}

\newcommand{\orcid}[1]{\href{https://orcid.org/#1}{\textcolor[HTML]{A6CE39}{\aiOrcid}}}

\newtheorem{lemma}{Lemma}
\newtheorem{proposition}{Proposition}
\newtheorem{corollary}{Corollary}

\newtheorem{definition}{Definition}

\definecolor{lime}{HTML}{A6CE39}
\DeclareRobustCommand{\orcidicon}{%
	\begin{tikzpicture}
	\draw[lime, fill=lime] (0,0)
	circle [radius=0.16]
	node[white] {{\fontfamily{qag}\selectfont \tiny ID}};
	\draw[white, fill=white] (-0.0625,0.095)
	circle [radius=0.007];
	\end{tikzpicture}
	\hspace{-2mm}
}

\foreach \x in {A, ..., Z}{%
	\expandafter\xdef\csname orcid\x\endcsname{\noexpand\href{https://orcid.org/\csname orcidauthor\x\endcsname}{\noexpand\orcidicon}}
}

\begin{document}
\title{Time-Encoding of Finite-Rate-of-Innovation Signals}
\author{\IEEEauthorblockN{
Abijith Jagannath Kamath,\orcidA{}~\IEEEmembership{Student Member,~IEEE},
Sunil Rudresh,\orcidB{} \\ and
Chandra Sekhar Seelamantula,\orcidC{}~\IEEEmembership{Senior Member,~IEEE}}

\thanks{
A.J. Kamath and C.S. Seelamantula are with the Department of Electrical Engineering (EE), Indian Institute of Science (IISc.), Bangalore (Email: \{abijithj, css\}@iisc.ac.in). S. Rudresh was with EE, IISc., where this work was carried out. He is presently with Walmart Global Tech., Bangalore (Email: sunilr.dvg@gmail.com).\\
\indent A part of Section \ref{sec:TEMFRI} of this paper was published in the Proceedings of IEEE International Conference on Acoustics, Speech, and Signal Processing (ICASSP), 2020. \\
\indent The figures in this paper are in colour in the electronic version.
}}

\markboth{Submitted to IEEE Transactions on Signal Processing}
{Kamath \MakeLowercase{\textit{et al.}}: Time-encoding of Finite-Rate-of-Innovation Signals}

\IEEEtitleabstractindextext{
\begin{abstract}
Time-encoding of continuous-time signals is an alternative sampling paradigm to conventional methods such as Shannon's sampling. In a time-encoding scheme, the signal is encoded using a sequence of time instants corresponding to an {\it \bfseries event}, and hence falls under {\it \bfseries event-driven sampling}. Time-encoding can be designed agnostic to the global clock of the sampling hardware, which makes the sampling asynchronous. Moreover, the encoding is sparse, which makes time-encoding energy efficient. However, the signal representation is nonstandard and in general, nonuniform. In this paper, we consider time-encoding of finite-rate-of-innovation signals, and in particular, periodic signals composed of weighted and time-shifted versions of a known pulse. We consider encoding using both crossing-time-encoding machine (C-TEM) and integrate-and-fire time-encoding machine (IF-TEM). By analyzing how time-encoding manifests in the Fourier domain, we arrive at the familiar {\it \bfseries sum-of-sinusoids} structure of the Fourier coefficients by means of a suitable linear transformation of the time-encoded measurements. Thereafter, standard FRI techniques become applicable. Further, we extend the theory to multichannel time-encoding such that each channel operates at a lower sampling requirement. We also study the effect of measurement noise, where the temporal measurements are perturbed by additive noise. To combat the effect of noise, we propose a robust optimization framework to simultaneously denoise the Fourier coefficients and estimate the annihilating filter accurately. We provide sufficient conditions for time-encoding and perfect reconstruction using C-TEM and IF-TEM, and provide simulation results to substantiate our findings.
\end{abstract}

\begin{IEEEkeywords}
Time-encoding machine (TEM), finite-rate-of-innovation (FRI) sampling, time-based sampling, crossing-time-encoding machine (C-TEM), integrate-and-fire time-encoding machine (IF-TEM).
\end{IEEEkeywords}}

 \setlength{\textfloatsep}{10pt} % Vertical space below (above) [t] ([b]) floats
 \setlength{\abovecaptionskip}{-2pt}
 \setlength{\belowcaptionskip}{0pt}
 \addtolength{\subfigcapskip}{-2pt}

\maketitle
\IEEEdisplaynontitleabstractindextext
\IEEEpeerreviewmaketitle

%% -----------------------------------------------------------
% 					INTRODUCTION
%% -----------------------------------------------------------

\section{Introduction}
\label{sec:Introduction}
\IEEEPARstart{S}{ampling} lies at the interface between analog signals and digital processing. Under certain conditions on the signal class and sampling rate, uniform samples of a signal or its filtered version constitute an accurate discrete representation with perfect reconstruction guarantees \cite{unser2000sampling}. The celebrated Shannon sampling framework \cite{shannon1949communication} is the gold standard for bandlimited signals and provides least-squares optimality guarantees while handling nonbandlimited signals. Time-encoding or time-based sampling is an alternative to conventional uniform sampling, where the discrete representation of the signal is a sequence of measurements along its abscissa. These measurements are determined by events, and in general, generate nonuniformly spaced time instants. The device used to obtain such measurements is called a {\it time-encoding machine} (TEM). Time-encoding methods have deep roots in neuroscience where representation of sensory information as a sequence of action potentials is encoded temporally \cite{adrian1928basis, burkitt2006review}. Lazar and T\'oth \cite{lazar2004period} designed the integrate-and-fire time-encoding machine (IF-TEM) that encodes time instants at which the running integral of the signal crosses a threshold. The sampling mechanism is asynchronous, implementable in real-time, energy efficient and generates sparse measurements. For example, when the signal is relatively constant, the time instants recorded are farther apart on the average as compared to signals that have a higher bandwidth. Therefore, time-encoding is intrinsically opportunistic. Such sampling mechanisms fall under the broad category of {\it event-driven sampling} or {\it neuromorphic sampling} and have paved the way for a new class of dynamic audio and vision sensors \cite{brandii2014sensor, delbruck2010sensors}.\\
\indent The discrete representation obtained using the IF-TEM is a sequence of nonuniform instants called the {\it trigger times}, together with the corresponding local averages. Such representations are forms of zero-crossing encoding and level-crossing encoding and have found several applications \cite{logan1977information,mallat1991zero,chandra2005instantaneous,marvasti2012nonuniform}. Signal reconstruction from trigger times is equivalent to the problem of reconstruction using nonuniform samples of a known transformation of the signal. When the signal is bandlimited, reconstruction from nonuniform temporal measurements and local averages is solved using an iterative method \cite{wiley1978recovery} based on Sandberg's theorem \cite{sandberg1963properties}. The iterations can be interpreted as projections on to convex sets (PoCS) \cite{karen2020multichannel}, alternating between maintaining measurement consistency and satisfying the bandlimitedness constraint. The iterations converge when the average sampling rate is greater than the Nyquist rate. Aldroubi and Gr\"ochenig \cite{aldroubi2001nonuniform} have extended the sampling theory to shift-invariant spaces. In this case, as with bandlimited signals, the reconstruction algorithm is of the PoCS type. The sampling requirement is of the order of the Nyquist rate \cite{gontier2014sampling}.\\
\indent In this paper, we consider time-encoding of signals that possess a finite rate of innovation (FRI) \cite{vetterli2002sampling}. The FRI model represents sparse analog signals for which the sampling requirement matches the rate of innovation. Consider a prototypical $T$-periodic FRI signal of the form
\begin{equation}
	x(t) = \sum_{m\in\zz}\sum_{k=0}^{K-1} c_{k} \varphi(t-\tau_{k}-mT),
\end{equation}
where $\varphi (t)$ is a known pulse and the parameters $\{(c_{k},\tau_{k})\}_{k=0}^{K-1}$ are not known. The rate of innovation of $x(t)$ is $\displaystyle \frac{2K}{T}$. If $\varphi (t)$ is a Dirac impulse, then $x(t)$ is not bandlimited. Yet, the rate of innovation and the requirement on the minimum number of measurements remain unchanged in the FRI framework, also referred to as {\it sub-Nyquist sampling}. The applications are in RADAR \cite{ilan2014radar, rudresh2017radar}, ultrasound \cite{tur2011innovation}, radioastronomy \cite{pan2016towards}, optical-coherence tomography \cite{mulleti2014fdoct}, etc.. One could also interpret FRI signals as belonging to a union of subspaces \cite{do2008union} where the parameters $\{\tau_{k}\}_{k=0}^{K-1}$ define the subspace. Hence, signal reconstruction requires localization of the subspace by estimating $\{\tau_{k}\}_{k=0}^{K-1}$ followed by estimation of the amplitudes $\{c_{k}\}_{k=0}^{K-1}$. We address reconstruction of FRI signals from measurements obtained using a C-TEM or an IF-TEM via the parameter estimation route and show that the sampling requirements for perfect reconstruction are of the order of the rate of innovation. The analysis also carries over to the multichannel setup, where each channel operates with a reduced sampling requirement. We also consider the presence of noise on the temporal measurements. To overcome the detrimental effects of noise, we deploy an alternating minimization strategy that jointly denoises the measurements and performs parameter estimation. The optimization strategy is similar to that of Generalized FRI (GenFRI) \cite{pan2016towards}. In view of the key ingredients, namely, TEM, FRI signals, and GenFRI for reconstruction, we refer to the proposed approach as GenFRI-TEM.

%% -----------------------------------------------------------

\subsection{Related Literature}
\label{subsec:RelatedLiterature}
Time-encoding of FRI signals is a new research direction. The problem was first posed by Alexandru and Dragotti \cite{alexandru2020tem}, who considered event-driven sampling of a stream of Dirac impulses using an exponential-reproducing kernel. They developed a novel {\it sequential reconstruction algorithm} that recovers one impulse at a time. Their  reconstruction strategy requires the support of the sampling kernel to be smaller than the spacing between two consecutive impulses. While this is convenient for reconstruction, it can be  restrictive as the choice of the sampling kernel now becomes signal-dependent. Satisfying the constraint also requires sampling beyond the rate of innovation. Hilton {\it et al.} \cite{hilton2021synaptic} use the sequential reconstruction algorithm on a stream of Dirac impulses filtered using an alpha synaptic function, such that the sampling requirement for perfect reconstruction can be guaranteed by tuning the parameters of the IF-TEM. We overcame the shortcomings of \cite{alexandru2020tem} by proposing a Fourier-domain reconstruction approach \cite{rudresh2020time}, which also forms the basis for this paper.\\
\indent Recently, Naaman {\it et al.} developed a hardware prototype and demonstrated time-encoding of FRI signals \cite{naaman2021temhardware}. Their system uses a counter to encode the trigger times, which constitute the output of the TEM. The trigger times are used to recover the FRI signal. While our manuscript was under preparation, we  came across the work of Naaman {\it et al.} \cite{naaman2021fritem}, who addressed the reconstruction problem in the case of IF-TEM in the presence of noise. Their method is similar to that of \cite{rudresh2020time} with the difference that their reconstruction strategy removes terms in the forward linear transformation that cause instability. However, the drawback is that the signal can be reconstructed up to a constant.

%% -----------------------------------------------------------

\subsection{Our Contribution}
We consider time-encoding of FRI signals and develop a Fourier-domain reconstruction strategy expanding on the idea presented in \cite{rudresh2020time} (Section~\ref{sec:preliminaries} and Section~\ref{sec:Problem}). We consider both flavours of time-encoding --- C-TEM as well as IF-TEM and develop a kernel based sampling and reconstruction strategy (Section~\ref{sec:TEMFRI}). Our strategy is less restrictive in the sense that, unlike \cite{alexandru2020tem}, the choice of the sampling kernel is not tightly coupled to the signal. The proposed methodology is applicable to an FRI signal that can be expressed in the form of a sum of weighted and time-shifted pulses. We also show how the developments can be adapted to accommodate multichannel time-encoding (Section~\ref{sec:MTEMFRI}). Further, we analyze the effect of measurement noise on the forward linear transformation and signal measurements and develop an optimization strategy for robust signal reconstruction (Section~\ref{sec:Noise_TEMFRI}). Experimental results demonstrate accurate reconstruction using the proposed technique (Section~\ref{sec:experiments}). The results for Dirac impulses are presented in the main document, whereas the results for streams of B-splines are given in the Supplementary Material.
%% -----------------------------------------------------------

\subsection{Notations}
We use the symbols $\rr^{\zz}$ and $\cc^{\zz}$ to respectively denote real-valued and complex-valued sequences defined on the integers, and $\rr^{\rr}$ to denote real-valued functions defined on the real line. The double square-bracket notation $\llbracket \ell, m \rrbracket$ stands for the set of integers $\{\ell, \ell+1, \cdots, m\}$.

%% -----------------------------------------------------------
% 		PRELIMINARIES
%% -----------------------------------------------------------

\section{Preliminaries}
\label{sec:preliminaries}
In this section, we recapitulate the functioning of time-encoding machines, in particular, crossing-time-encoding machine (C-TEM) and integrate-and-fire time-encoding machine (IF-TEM). We also recall the Toeplitzification operator and Prony's method, which are central to FRI signal processing.

%% -----------------------------------------------------------
\subsection{Time-Encoding Machines and Sampling Sets}
\label{subsec:prelim_tem}
A TEM maps a function space $\rr^{\rr}$ to a real-valued sequence of time instants $\rr^{\zz}$. A formal definition follows along the lines put forth by Gontier and Vetterli \cite{gontier2014sampling}.
\begin{definition}
A time-encoding machine with an event operator $\sE:\rr^{\rr} \rightarrow \rr^{\rr}$ and references $\{r_{i} \in \rr^{\rr}\}_{i\in\zz}$ is a map $\sT: \rr^{\rr}\rightarrow \rr^{\zz}$ such that $\rr^{\rr} \ni y \mapsto \sT y$, with
\begin{enumerate}[label=\alph*.]
	\item $\sT y = \{t_i \in \rr \; | \; t_i > t_j, \; \forall i>j, i \in \zz\}$,
	\item $\displaystyle \lim_{n\rightarrow\pm\infty} t_{n} = \pm \infty$,\, \text{and}
	\item $(\sE y)(t_i) = r_{i}(t_{i}), \forall t_{i} \in \sT y$.
\end{enumerate}
\label{def:tem}
\end{definition}
A TEM outputs a sampling set of ``trigger times'' $\sT y$ with strictly increasing entries, which encodes the input signal $y(t)$. Uniform sampling can be viewed as a special case with $\sE$ as the identity operator and the references satisfying $r_{i}(t) = t - (t_{i-1}+T_{s})+y(t_{i})$, where $T_{s}$ is the sampling interval.
\begin{definition} The sampling density $d$ of the sampling set $\mathcal{X} = \{t_i\}_{i\in\zz}$ is defined as
    \begin{equation}
    	d(\mathcal{X}) = \sup_{i\in\zz} | t_{i+1} - t_i |.
    \end{equation}
\label{def:samplingDensity}
\end{definition}

\begin{definition} A sampling set $\mathcal{X} = \{t_i\}_{i\in\zz}$ is said to be $\epsilon$-distinct if $| t_{i+1} - t_{i} | > \epsilon, \; \forall t_{i} \in \mathcal{X}$.
\end{definition}

In the case of uniform sampling, the sampling density is  the sampling interval $T_{s}$, and the sampling set is $\epsilon$-distinct for $\epsilon < T_{s}$. We are interested in TEMs that have a bounded sampling density and an $\epsilon$-distinct sampling set. Although a TEM records only the trigger times, we obtain, via the event operator, samples of the transformed signal $(\sE y)(\sT y)$, which can be computed using the references $\{r_{i}\}_{i\in\zz}$ (cf. Definition~\ref{def:tem}). Hence, the TEM explicitly provides the sampling set $\sT y$ along with $(\sE y)(\sT y)$ in the absence of noise.
%% -----------------------------------------------------------

\subsection{Crossing-Time-Encoding Machine}
\label{subsubsec:prelim_ctem}
\begin{figure}[t]
\centering
\includegraphics[width=3in]{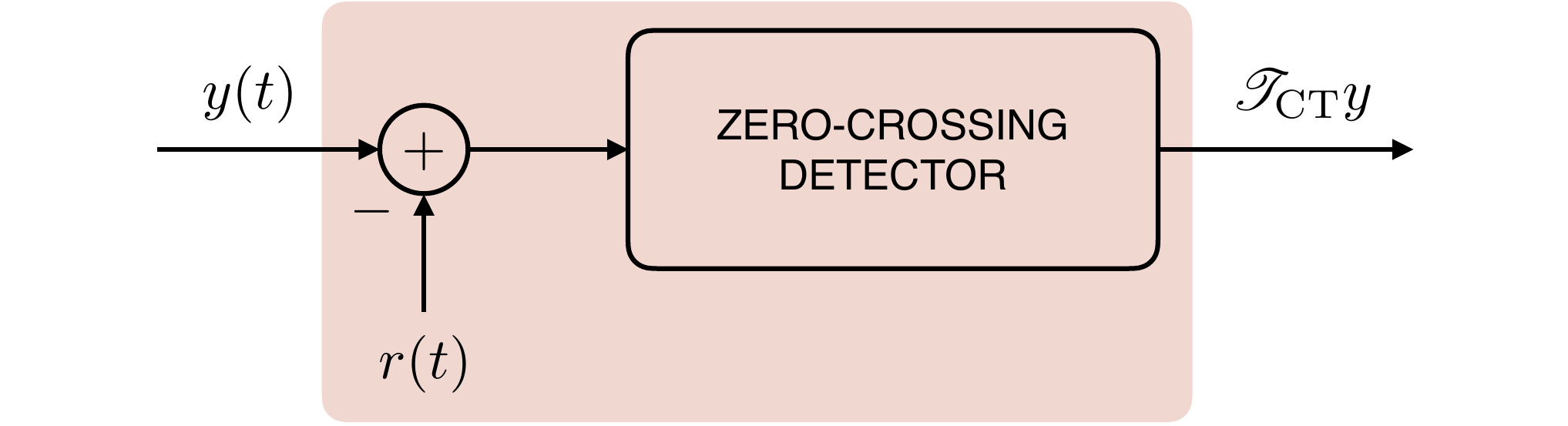}
\caption{The schematic of a crossing-time-encoding machine (C-TEM) with input $y(t)$ and reference $r(t)$. The zero-crossing detector outputs the instants at which $y(t)-r(t)=0$.}
\label{fig:schematicCTEM}
\end{figure}

The crossing-time-encoding machine (C-TEM) is a generalized zero-crossing detector. The signal to be encoded is matched with a reference sinusoidal signal and represented by the time instants where the signal and the reference match exactly, i.e., their difference equals zero. Figure~\ref{fig:schematicCTEM} shows the implementation of the C-TEM using a zero-crossing detector. Let $y(t)$ be the input to the C-TEM with reference $r(t)$. The output of the C-TEM is the sampling set $\sT_{\mathrm{CT}}y = \{t_{n} \; \vert \; y(t_{n}) = r(t_{n}), n\in \zz\}$. Logan \cite{logan1977information} addressed the reconstruction of bandpass signals from their zero-crossings. Zero-crossing detectors are robust to amplitude clipping that occurs when the signal goes outside the dynamic range of the sampling device. Such encodings are nonlinear and not invertible. Bar-David \cite{bar1974implicit} showed that the zero-crossings of the difference between a bandlimited signal and a sine-wave of a sufficient frequency is an invertible encoding of the bandlimited signal in the sense that it is possible to retrieve the bandlimited signals from the time-encoded measurements. Such an encoding can be interpreted as the output of a time-encoding machine where the signal is matched with a sinusoid. Employing a sinusoidal reference has the advantage that it ensures a bounded sampling density under mild conditions on the amplitude of the sinusoid.
Consider a C-TEM $\sT_{\mathrm{CT}}$ with sinusoidal reference $r(t) = A_{r}\cos(2\pi f_{r}t + \phi_{r})$, and identity event operator, i.e., $\sE = \mathrm{Id}$. Using Definition~\ref{def:tem}, $\sT_{\mathrm{CT}}y$ can be used to construct the sequence $\{y(t_{n})\}_{n\in\zz}$ using $r(t)$. The corresponding sampling density is derived next.

\begin{lemma} Let $y(t)$ the input to a C-TEM with reference $r(t) = A_{r}\cos(2\pi f_{r}t)$. Suppose $A_{r} \geq \Vert y \Vert_{\infty}$, the output of the C-TEM $\sT_{\mathrm{CT}}y = \{t_{n} \; \vert \; y(t_{n}) = r(t_{n}), n\in \zz\}$ satisfies $\displaystyle d(\sT_{\mathrm{CT}}y) = \sup_{n\in\zz} \vert t_{n+1} - t_{n} \vert < \frac{1}{f_{r}}$.
    \begin{proof}
	    See Appendix~\ref{appendix:proofLemmactemSamplingDensity}.
    \end{proof}
\label{lem:ctemSamplingDensity}
\end{lemma}
%% -----------------------------------------------------------

\subsection{Integrate-and-Fire Time-Encoding Machine}
\label{subsubsec:prelim_iftem}
\begin{figure}[t]
\centering
\includegraphics[width=3in]{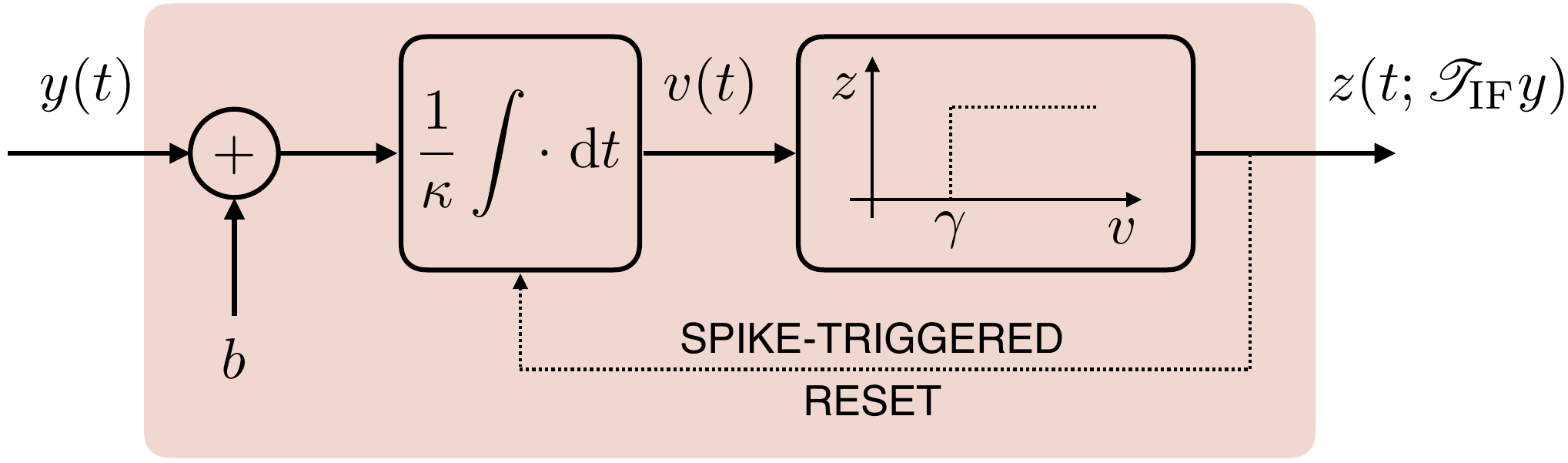}
\caption{The schematic of an integrate-and-fire time-encoding machine (IF-TEM) with parameters $\{b,\kappa,\gamma\}$. The integrator is reset every time a spike occurs at the output.}
\label{fig:schematicIFTEM}
\end{figure}

The integrate-and-fire time-encoding machine (IF-TEM) is inspired by neural encoding of sensory stimuli. Neural encoding has a refractory period, but we use a simplified IF-TEM model without the refractory period taken into account \cite{lazar2004period} (cf. Figure~\ref{fig:schematicIFTEM}). The input is offset by a constant $b>0$ and then integrated and scaled by $\kappa$. The result $v(t)$ is compared against a threshold $\gamma$. Whenever $v(t)$ equals $\gamma$, a spike is generated, which resets the integrator causing the output $v(t)$ to go to zero and start building up all over again. The output is a bilevel signal $z(\cdot;\sT_{\mathrm{IF}}y)$ with transitions at $t_n\in\sT_{\mathrm{IF}}y$. Extracting the sampling set $\sT_{\mathrm{IF}}y$ from $z(\cdot;\sT_{\mathrm{IF}}y)$ is possible using the B-spline method described in \cite{vetterli2002sampling} or the E-spline method described in \cite{seelamantula2010sub}. For the ensuing discussion, we assume that the sampling set $\sT_{\mathrm{IF}}y = \{t_{n}\}_{n\in\zz}$ is available as the output of the IF-TEM. The trigger times are related to the local averages of the input $y(t)$ as shown in \cite{lazar2004period}, recalled below.

\begin{lemma} Let $y(t)$ be the input to an IF-TEM (Figure~\ref{fig:schematicIFTEM}) with parameters $\{b,\kappa,\gamma\}$. The output of the IF-TEM is a set of strictly increasing instants $\sT_{\mathrm{IF}}y = \{t_{n}\}_{n\in \zz}$ such that
\begin{equation}
\bar{y}_{n} = \int_{t_{n}}^{t_{n+1}} y(t)\,\, \dd t = -b(t_{n+1} - t_{n}) + \kappa \gamma, \forall n\in \zz.
\end{equation}
\label{lem:ttransform}
\end{lemma}
The proof is available in \cite{lazar2004period}. Lemma~\ref{lem:ttransform} defines the event operator of the IF-TEM as an accumulator, which aids in the construction of the sequence $\{\bar{y}_{n}\}_{n\in\zz}$, given the output of the IF-TEM $\sT_{\mathrm{IF}}y$. The IF-TEM generates a bounded and $\epsilon$-distinct sampling set, as shown next.
\begin{corollary}
Let $y(t)$ be the input to an IF-TEM (Figure~\ref{fig:schematicIFTEM}) with parameters $\{b,\kappa,\gamma\}$, with $\Vert y \Vert_{\infty} < b$. The output $\sT_{\mathrm{IF}}y = \{t_{n}\}_{n\in\zz}$ satisfies:
\begin{equation}
	\frac{\kappa\gamma}{b+\Vert y \Vert_{\infty}} \leq t_{n+1} - t_{n} \leq \frac{\kappa\gamma}{b-\Vert y \Vert_{\infty}}.
\end{equation}
\label{cor:iftemSamplingDensity}
\end{corollary}
The proof is provided in \cite{lazar2004period}. Corollary~\ref{cor:iftemSamplingDensity} places an upper bound on the sampling density for a bounded input. The sampling sets are $\epsilon$-distinct for $\displaystyle \epsilon < \frac{\kappa\gamma}{b+\Vert y \Vert_{\infty}}$.

\subsection{Toeplitzification Operator}
\label{subsec:prelim_toeplitzification}
Consider the vectors $\bx = [x_{-M} \; \cdots \; x_M]^\TT\in\cc^{N}$ and $\bu=[u_{1} \; \cdots \; u_{P+1}]^\TT\in\cc^{P+1}$. The convolution of the sequences $\tilde{\boldsymbol{x}} = [\cdots \; 0 \; x_{-M} \; \cdots \; \boxed{x_0} \; \cdots \; x_M \; 0 \; \cdots] \in \cc^{\zz}$ and $\tilde{\boldsymbol{u}} = [\cdots \; \boxed{0} \; u_1 \; \cdots \; u_{P+1} \; 0 \; \cdots]\in\cc^{\zz}$ can be expressed as the product of a Toeplitz matrix constructed using $\bx$ and the vector $\bu$ or vice versa.
\begin{definition} Let $\bx \in \cc^N$, $N=2M+1$, for some $M\in\nn$ with entries indexed as $\bx = [x_{-M}\; x_{-M+1} \; \cdots \; x_{M-1}\; x_M]^\TT$, and, for any $P\leq M$, consider the set of Toeplitz matrices $\mathbb{T}_P\subset \cc^{(N-P)\times (P+1)}$. Then, $\bx$ can be embedded into a Toeplitz matrix in $\mathbb{T}_{P}$, using the Toeplitzification operator, $\Gamma_P: \cc^N \rightarrow \mathbb{T}_P$, with $\bx \mapsto (\Gamma_P\bx)_{i,j} = x_{-M+P+i-j},\; i=1,\cdots,N-P,\; j=1,\cdots,P+1$.
\label{def:toep}
\end{definition}
The vector $\bx$ is called the {\it generator} of the matrix $\Gamma_P\bx$. $(\Gamma_P\bx)\bu$ denotes the valid part of the convolution between $\tilde{\boldsymbol{x}}$ and $\tilde{\boldsymbol{u}}$. For $P=M$, there exists an operator $\zeta_{N}:\cc^{P+1}\mapsto\cc^{(P+1)\times N}$, which is the right dual of $\Gamma_{M}$, such that $(\Gamma_{M}\bx)\bu = (\zeta_{N}\bu)\bx, \forall \bx, \bu$. The right dual relies on the commutativity of the convolution operation: $\tilde{\boldsymbol{x}} * \tilde{\boldsymbol{u}} = \tilde{\boldsymbol{u}} * \tilde{\boldsymbol{x}}$.

%% -----------------------------------------------------------
\subsection{Prony's Method}
\label{subsec:pronysmethod}
Consider a $T$-periodic stream of Dirac impulses
\begin{equation*} 
	x(t) = \sum_{m\in\zz}\sum_{k=0}^{K-1} c_{k} \delta(t-\tau_{k} - mT).
\end{equation*}
Using Poisson summation formula, $x(t)$ can be expressed as
\begin{equation}
	x(t) = \sum_{m\in\zz} \underbrace{\frac{1}{T}\sum_{k=0}^{K-1} c_{k} e^{-\jj 2\pi m \tau_{k}/T}}_{\hat{x}_{m}} e^{\jj 2\pi m t/T}.
\label{eq:swce_dirac}
\end{equation}
The Fourier coefficients $\hat{\boldsymbol{x}} = \{\hat{x}_m\}_{m\in\zz} \in \cc^{\zz}$ are in the sum-of-weighted-complex-exponentials (SWCE) form. The problem is to estimate the parameters $\{\tau_{k}\}_{k=0}^{K-1}$ from $\hat{\boldsymbol{x}}$. This is possible using high-resolution spectral estimation (HRSE) techniques. In particular, Prony's method \cite{prony} requires at least $2K+1$ contiguous Fourier coefficients and employs a $(K+1)$-tap annihilating filter $\boldsymbol{h} = [\cdots \; 0 \; h_0 \; h_1 \; \cdots \; h_{K} \; 0 \; \cdots]\in\cc^{\zz}$ with roots $\{\vartheta_k = e^{-\jj 2\pi \tau_k/T}\}_{k=0}^{K-1}$, which are in one-to-one correspondence with the parameters $\{\tau_{k}\}_{k=0}^{K-1}$. The $\mathcal{Z}$-transform of the annihilating filter is given by
\begin{equation*}
	H(z) = \sum_{k=0}^{K} h_{k} z^{-k} = h_{0}\prod_{k=0}^{K-1}(1-\vartheta_{k} z^{-1}).
\end{equation*}
The {\it annihilation property} is easy to verify:
\begin{equation}
\begin{split}
	(\boldsymbol{h}*\hat{\boldsymbol{x}})_m &= \sum_{k\in\zz} h_{k} \hat{x}_{m-k} = \sum_{k=0}^{K} h_{k}  \frac{1}{T}\sum_{k'=0}^{K-1} c_{k'} \vartheta_{k'}^{(m-k)}, \\
	&= \frac{1}{T} \sum_{k'=0}^{K-1} c_{k'} \vartheta_{k'}^{m}  \underbrace{\sum_{k=0}^{K} h_{k} \vartheta_{k'}^{-k}}_{H(\vartheta_{k'})=0} = 0, \; \forall m\in\zz.
\end{split}
\label{eq:annihilation}
\end{equation}
Considering $m \in \llbracket -M+K, M \rrbracket$, $M\geq K$, gives the linear system of equations $(\Gamma_{K}\hat{\bx})\bh = \zerovec$, where $\hat{\bx} = [\hat{x}_{-M} \; \cdots \; \hat{x}_{M}]^{\TT}\in\cc^{N}$ is the vector of Fourier coefficients, and $\bh = [h_{0} \; \cdots \; h_{K}]^{\TT}\in\cc^{K+1}$ constitutes the annihilating filter coefficients. It can be shown that $\rank(\Gamma_{K}\hat{\bx}) = K$. The annihilating filter is a nontrivial vector in the null space of $\Gamma_{K}\hat{\bx}$ and can be found using the Eckart-Young theorem \cite{horn2012matrix}, which selects the right eigenvector corresponding to the smallest singular value of the matrix $\Gamma_K\hat{\bx}$.

%% -----------------------------------------------------------
% 				PROBLEM FORMULATION
%% -----------------------------------------------------------

\section{Problem Formulation}
\label{sec:Problem}
\begin{figure}[t]
\centering
\includegraphics[width=3.5in]{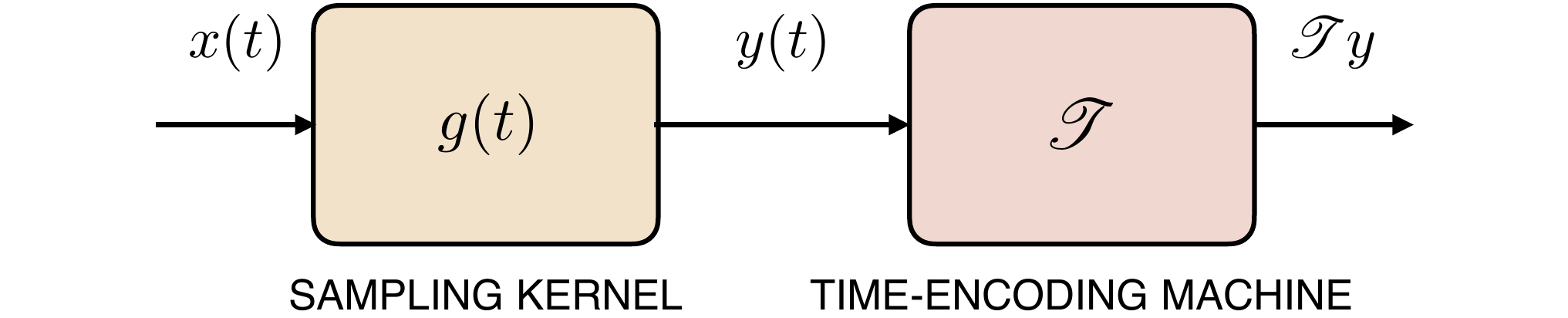}
\caption{A schematic of kernel-based time-encoding proposed in this paper. The sampling kernel is designed to satisfy the {\it alias-cancellation} constraints (Eq.~\eqref{eq:samplingKernel}), which facilitates signal reconstruction from the trigger times using a Fourier-domain approach. The time-encoding machine could be either a C-TEM or an IF-TEM.}
\label{fig:schematicKernelBasedSampling}
\end{figure}

Consider a $T$-periodic signal $x \in L^{2}([0,T[)$ obtained by a linear combination of delayed versions of a prototype pulse $\varphi \in L^{2}(\rr)$:
\begin{equation}
x(t) = \sum_{m\in\zz}\sum_{k=0}^{K-1} c_{k} \varphi(t-\tau_{k} - mT),
\label{eq:signalModel}
\end{equation}
where $\{c_{k}\in\rr\}_{k=0}^{K-1}$ are the unknown amplitudes and $\{\tau_{k} \in [0,T[\}_{k=0}^{K-1}$ are the unknown shifts. Clearly, $x(t)$ has a rate of innovation of $\displaystyle \frac{2K}{T}$ and the minimum sampling requirement is $2K$ samples per period. Since $x\in L^{2}([0,T[)$, we could express it using Fourier series:
\begin{equation}
	x(t) = \sum_{m\in\zz} \hat{x}_{m} e^{\jj \omega_{0}m t}, \; \omega_{0} = \frac{2\pi}{T}.
\end{equation}
Using Eq.~\eqref{eq:signalModel} and Poisson summation formula, the Fourier coefficients are given by
\begin{equation}
	\hat{x}_{m} = \frac{1}{T}\hat{\varphi}(m\omega_{0})\sum_{k=0}^{K-1} c_{k} e^{-\jj \omega_{0}m \tau_{k}},
\label{eq:swce_pulse}
\end{equation}
where $\hat{\varphi}$ is the Fourier transform of $\varphi$. The Fourier coefficients have the SWCE form. Estimation of the shifts from the Fourier coefficients is achieved with $2K+1$ contiguous samples of $\hat{\boldsymbol{x}}$ using Prony's method.\\
\indent We consider kernel-based time-encoding (cf. Figure~\ref{fig:schematicKernelBasedSampling}) of the $T$-periodic FRI signal $x(t)$, using a suitable sampling kernel $g(t)$ and a TEM to give the sampling set $\sT y$ and the corresponding signal measurements $(\sE y)(\sT y)$, where $y(t)=(x*g)(t)$. The reconstruction problem is posed as follows: Given $\sT y$, reconstruct the FRI signal in  Eq.~\eqref{eq:signalModel} or equivalently, determine the amplitudes and shifts. \\
\indent The analysis extends to aperiodic FRI signals of the type
\begin{equation*}
	x_{ap}(t) = \sum_{k=0}^{K-1}c_k \varphi(t-\tau_k),
\end{equation*}
by employing the periodized sampling kernel $\tilde{g}(t) = \sum_{m\in\zz}g(t-mT)$. It is easy to show that $y(t) = (x*g)(t) = (x_{ap}*\tilde{g})(t)$. The analysis for aperiodic FRI signals follows along the lines of periodic FRI signals \cite{rudresh2020time}.

%% -----------------------------------------------------------
%   TIME-ENCODING OF FRI SIGNALS
%% -----------------------------------------------------------
\section{Kernel Based Sampling and Reconstruction}
\label{sec:TEMFRI}
We propose kernel-based time-encoding of the FRI signal in Eq.~\eqref{eq:signalModel} using a suitable sampling kernel such that it becomes possible to determine the Fourier coefficients by solving a linear system of equations. Filtering the FRI signal $x(t)$ using a sampling kernel $g(t)$ yields the output
\begin{equation}
\begin{split}
	y(t) &= (x*g)(t) = \int_{\rr} g(\nu) x(t-\nu) \,\dd \nu, \\
	&= \int_{\rr} g(\nu) \sum_{m\in\zz} \hat{x}_{m} e^{\jj \omega_{0}m (t-\nu)}\, \dd \nu, \\
	&= \sum_{m\in\zz} \hat{x}_{m} \left( \int_{\rr}g(\nu) e^{-\jj \omega_{0}m \nu}\, \dd \nu\right) e^{\jj \omega_{0}m t}, \\
	&= \sum_{m\in\zz} \hat{x}_{m} \hat{g}(m\omega_{0})e^{\jj \omega_{0}m t},
\end{split}
\label{eq:filteredSignalDerivation}
\end{equation}
where $\hat{g}$ denotes the Fourier transform of $g$. The filtered signal $y(t)$ is also $T$-periodic with Fourier coefficients $\{\hat{x}_{m} \hat{g}(m\omega_{0})\}_{m\in\zz}$. Assume that the sampling kernel satisfies Fourier-domain alias-cancellation conditions \cite{mulleti2017paley}:
\begin{equation}
\hat{g}(m\omega_{0}) = \begin{cases}
g_{m} \neq 0, & m \in \llbracket -M, M \rrbracket, \\
0, & m \notin \llbracket -M, M \rrbracket. \\
\end{cases}
\label{eq:samplingKernel}
\end{equation}
Examples of such kernels include the sinc function of bandwidth $\displaystyle \frac{2M+1}{T}$, where $g_{m}=1, \forall m\in\llbracket -M, M \rrbracket$; the sum-of-sincs (SoS) kernel in the Fourier domain \cite{tur2011innovation}; the sum-of-modulated-spline (SMS) kernels in the time domain \cite{mulleti2017paley}; and exponential-reproducing and polynomial-reproducing kernels, which satisfy the generalized Strang-Fix conditions \cite{dragotti2007fix}. Setting $g_{m}=1, \forall m\in\llbracket -M, M \rrbracket$ in Eq.~\eqref{eq:samplingKernel}, without loss of generality, we have
\begin{equation}
y(t) = (x*g)(t) = \sum_{m=-M}^{M} \hat{x}_{m} e^{\jj \omega_{0}m t}.
\label{eq:filteredSignal}
\end{equation}
Effectively, we have a trigonometric polynomial from which it is possible to construct a linear system of equations in the Fourier coefficients $\hat{\boldsymbol{x}}$ using $\sT y$ and $(\sE y)(\sT y)$. We investigate the sampling and reconstruction aspects, first using a C-TEM and then using an IF-TEM, to show that, in both cases, the Fourier coefficients are related to the measurements by a linear transformation.
%% -----------------------------------------------------------

\subsection{Crossing-Time-Encoding of FRI Signals}
\label{sec:CTEM_sampling}
Let $\sT_{\mathrm{CT}}y = \{t_{n} \; \vert \; y(t_{n}) = r(t_{n}), n\in \zz\}$ be the output of the C-TEM with $y(t) = (x*g)(t) $ as the input and a sinusoidal reference $r(t) = A_{r}\cos(2\pi f_{r}t)$. We have the  measurements
\begin{equation}
y(t_{n}) = \sum_{m=-M}^{M} \hat{x}_{m} e^{\jj \omega_{0}m t_{n}}, \forall n\in\zz.
\label{eq:CTEMmeasurements}
\end{equation}
Considering $L$ measurements in the interval $[0,T[$ gives rise to a linear system of equations $\by = \bG_{\mathrm{CT}}\hat{\bx}$, where $\by = [y(t_{1}) \; \cdots \; y(t_{L})]^{\TT}\in\rr^{L}$ and $\bG_{\mathrm{CT}}\in\cc^{L\times N}$ is given by
\begin{equation}
	\bG_{\mathrm{CT}} =
    \begin{bmatrix}
    e^{-\mathrm{j}M\omega_0 t_{1}}  & \hspace{-0.2cm}\cdots & \hspace{-0.1cm}e^{-\mathrm{j}\omega_0 t_{1}} & 1  & e^{\mathrm{j}\omega_0 t_{1}} &\hspace{-0.2cm}\cdots &\hspace{-0.1cm}e^{\mathrm{j}M\omega_0 t_{1}}\\
    e^{-\mathrm{j}M\omega_0 t_{2}}  & \hspace{-0.2cm}\cdots &\hspace{-0.1cm} e^{-\mathrm{j}\omega_0 t_{2}} & 1 & e^{\mathrm{j}\omega_0 t_{2}} &\hspace{-0.2cm}\cdots &\hspace{-0.1cm}e^{\mathrm{j}M\omega_0 t_{2}}\\
    \vdots  &  & \vdots & \vdots & \vdots & &\hspace{-0.1cm}\vdots\\
    e^{-\mathrm{j}M\omega_0 t_{L}}  & \hspace{-0.2cm}\cdots & \hspace{-0.1cm}e^{-\mathrm{j}\omega_0 t_{L}} & 1 & e^{\mathrm{j}\omega_0 t_{L}} &\hspace{-0.2cm}\cdots &\hspace{-0.1cm}e^{\mathrm{j}M\omega_0 t_{L}}  \/
    \end{bmatrix}.
\label{eq:ctemMatrix}
\end{equation}
\\
\begin{lemma} The matrix $\bG_{\mathrm{CT}} \in \cc^{L\times N}$ defined in Eq.~\eqref{eq:ctemMatrix} has full column-rank whenever $L \geq N$.
    \begin{proof}
    See Appendix~\ref{appendix:proofLemmactemMatrix}.
    \end{proof}
\label{lem:ctemMatrix}
\end{lemma}
% $\bG_{\mathrm{CT}}$ is the forward linear transformation under uniform and random sampling --- and $\bG_{\mathrm{CT}}$ is left-invertible when the sampling set is $\epsilon$-distinct for some $\epsilon>0$ (cf. Appendix~\ref{appendix:proofLemmactemMatrix}). This is consequential of the fact that the C-TEM may be viewed as a structured event-driven method to obtain nonuniform samples. \\
\indent By virtue of Lemma~\ref{lem:ctemMatrix}, for $L\geq N$, the matrix $\bG_{\mathrm{CT}}$ is left-invertible and the solution to the linear system $\by = \bG_{\mathrm{CT}}\hat{\bx}$ is unique. The shifts can be determined using Prony's method and the amplitudes can be determined by linear regression (cf. Eq.~\eqref{eq:swce_pulse}). The following result gives sufficient conditions for perfect reconstruction of Eq.~\eqref{eq:signalModel} from C-TEM measurements.
\begin{proposition}
Consider crossing-time-encoding of the $T$-periodic FRI signal $x(t)$ in Eq.~\eqref{eq:signalModel} with reference $r(t)$ and a sampling kernel $g(t)$ that satisfies the alias-cancellation conditions (Eq.~\eqref{eq:samplingKernel}). The set of time instants $\{t_{n}\}_{n=1}^{L} \subset \sT_{\mathrm{CT}}(x*g) \cap [0,T[$ is a sufficient representation of $x(t)$ when $L\geq 2K+1$ and the reference is the sinusoid $r(t) = A_{r}\cos(2\pi f_{r}t)$ with $A_{r} > \Vert x*g \Vert_{\infty}$ and $\displaystyle f_{r}\geq \frac{2K+1}{T}$.
  \begin{proof}
    See Appendix~\ref{appendix:proofPropCtemGuarantee}.
  \end{proof}
\label{prop:ctemGuarantees}
\end{proposition}
%% -----------------------------------------------------------

\subsection{Integrate-and-Fire Time-Encoding of FRI Signals}
\label{sec:IFTEM_sampling}
Let $\sT_{\mathrm{IF}}y = \{t_{n}\}_{n\in \zz}$ be the output of the IF-TEM with parameters $\{b,\kappa,\gamma\}$ and the filtered signal $y(t)=(x*g)(t)$ as the input. Using Lemma~\ref{lem:ttransform}, the measurements take the form:
\begin{equation}
\begin{split}
	\bar{y}_{n} = \sum_{\substack{m=-M \\ m \neq 0}}^{M} & \frac{\hat{x}_{m}}{\jj \omega_{0}m} \left( e^{\jj \omega_{0}m t_{n+1}} - e^{\jj \omega_{0}m t_{n}} \right) \\
	& + \; \hat{x}_{0} \left( t_{n+1} - t_{n} \right), \; \forall n\in \zz.
\end{split}
\label{eq:iftemMeasurements}
\end{equation}
Considering $L$ measurements in the interval $[0,T[$, we obtain a linear system of equations of the form $\bar{\by} = \bG_{\mathrm{IF}}\hat{\bx}$, where $\bar{\by} = [\bar{y}_{1} \; \bar{y}_{2} \; \cdots \; \bar{y}_{L-1}]^{\TT}\in\rr^{L-1}$ and $\bG_{\mathrm{IF}}\in\cc^{(L-1)\times N}$ is as defined in Eq.~\eqref{eq:iftemMatrix}.
\begin{lemma}
The matrix $\bG_{\mathrm{IF}}$ defined in Eq.~\eqref{eq:iftemMatrix} has full column-rank whenever $L \geq N+1$.
    \begin{proof}
	    See Appendix~\ref{appendix:proofLemmaiftemMatrix}.
    \end{proof}
\label{lem:iftemMatrix}
\end{lemma}
Using Lemma~\ref{lem:iftemMatrix}, with $L\geq N+1$, the matrix $\bG_{\mathrm{IF}}$ is left-invertible and the linear system $\bar{\by} = \bG_{\mathrm{IF}}\hat{\bx}$ has a unique solution. Prony's method can be used to determine the shifts and the amplitudes can be estimated using linear regression. The following result gives sufficient conditions for perfect reconstruction of $x(t)$ in Eq.~\eqref{eq:signalModel} from IF-TEM measurements.
\begin{proposition}
Consider integrate-and-fire time-encoding of the $T$-periodic FRI signal $x(t)$ in Eq.~\eqref{eq:signalModel} with parameters $\{b,\kappa,\gamma\}$ (cf. Figure~\ref{fig:schematicIFTEM}) and a sampling kernel $g(t)$ that satisfies the alias-cancellation conditions given in Eq.~\eqref{eq:samplingKernel}. The set of time instants $\{t_{n}\}_{n=1}^{L} \subset \sT_{\mathrm{IF}}(x*g) \cap [0,T[$ constitutes a sufficient representation of $x(t)$ with $(L-1)\geq (2K+1)$ if the parameters of the IF-TEM satisfy the following condition:
\begin{equation*}
	\frac{\kappa\gamma}{b-\Vert x*g \Vert_{\infty}} < \frac{T}{L}.
\end{equation*}
    \begin{proof}
    	See Appendix~\ref{appendix:proofPropIFtemGuarantee}.
    \end{proof}
\label{prop:iftemGuarantees}
\end{proposition}
%\begin{figure*}[t]
\begin{strip}
\vspace{1em}
% \hrulefill
\begin{equation}
	\bG_{\mathrm{IF}} =
    \begin{bmatrix}
    e^{-\mathrm{j}M\omega_0 t_{2}}-e^{-\mathrm{j}M\omega_0 t_{1}}  & \cdots & e^{-\mathrm{j}\omega_0 t_{2}}-e^{-\mathrm{j}\omega_0 t_{1}} & t_{2}-t_{1} & e^{\mathrm{j}\omega_0 t_{2}}-e^{\mathrm{j}\omega_0 t_{1}} & \cdots & e^{\mathrm{j}M\omega_0 t_{2}}-e^{\mathrm{j}M\omega_0 t_{1}}\\
    e^{-\mathrm{j}M\omega_0 t_{3}}-e^{-\mathrm{j}M\omega_0 t_{2}}  & \cdots & e^{-\mathrm{j}\omega_0 t_{3}}-e^{-\mathrm{j}\omega_0 t_{2}} &  t_{3}-t_{2} & e^{\mathrm{j}\omega_0 t_{3}}-e^{\mathrm{j}\omega_0 t_{2}} & \cdots & e^{\mathrm{j}M\omega_0 t_{3}}-e^{\mathrm{j}M\omega_0 t_{2}}\\
    \vdots  & \ddots & \vdots & \vdots & \vdots & \ddots & \vdots\\
    e^{-\mathrm{j}M\omega_0 t_{L}}-e^{-\mathrm{j}M\omega_0 t_{L-1}} & \cdots & e^{-\mathrm{j}\omega_0 t_{L}}-e^{-\mathrm{j}\omega_0 t_{L-1}} & t_{L}-t_{L-1} & e^{\mathrm{j}\omega_0 t_{L}}-e^{\mathrm{j}\omega_0 t_{L-1}} & \cdots & e^{\mathrm{j}M\omega_0 t_{L}}-e^{\mathrm{j}M\omega_0 t_{L-1}} \/
    \end{bmatrix}
\label{eq:iftemMatrix}
\end{equation}
\vspace{0.5em}
% \hrulefill
\end{strip}
%\end{figure*}

%% -----------------------------------------------------------
% 				MULTICHANNEL SAMPLING
%% -----------------------------------------------------------

\section{Multichannel Time-Encoding}
\label{sec:MTEMFRI}
\begin{figure}[t]
\centering
\includegraphics[width=3.5in]{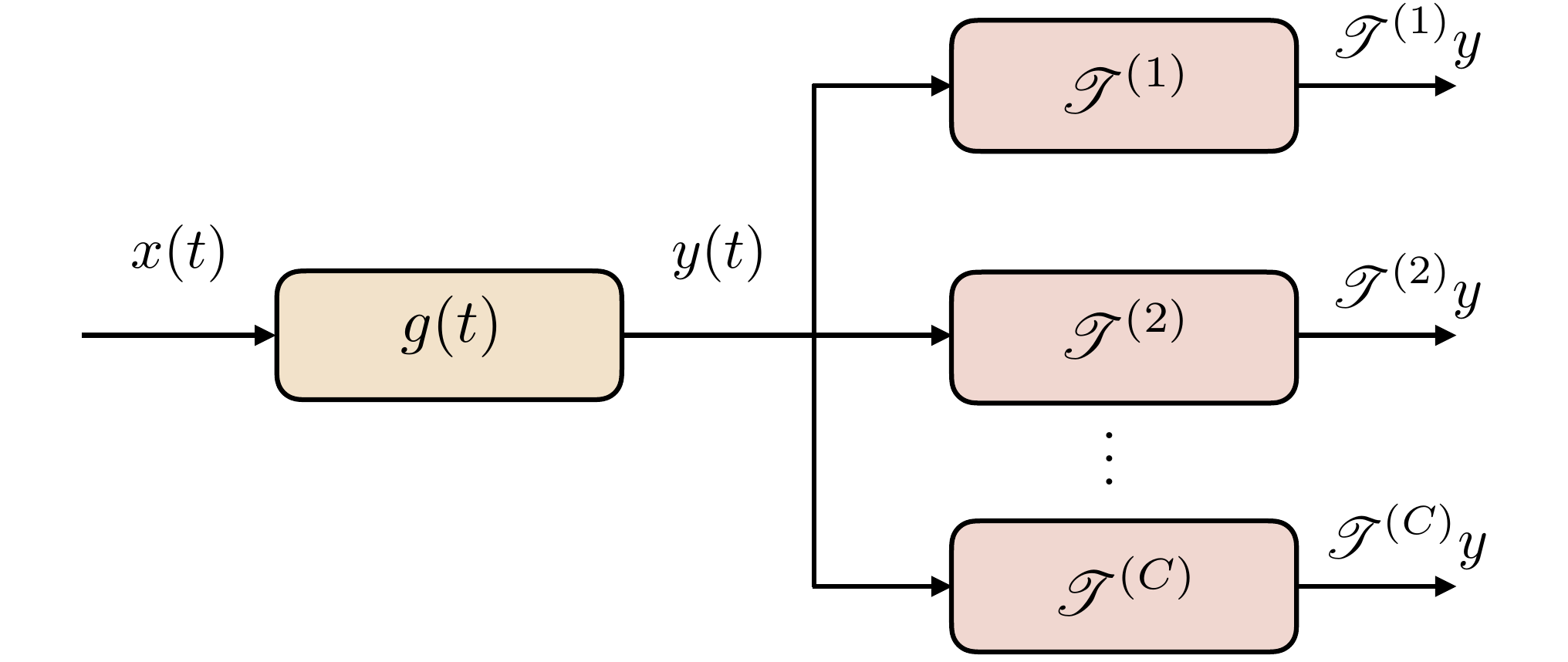}
\caption{A schematic of kernel-based multichannel time-encoding using a sampling kernel $g(t)$ and time-encoding machines $\sT^{(1)}, \sT^{(2)} \cdots, \sT^{(C)}$. The sampling kernel is designed to satisfy the alias-cancellation conditions.}
\label{fig:schematicKernelBasedMultichannelSampling}
\end{figure}

In this section, we analyze multichannel time-encoding of FRI signals. The sufficient condition in Proposition~\ref{prop:ctemGuarantees} may stipulate samples more than the rate of innovation of the signal. The sampling requirement may be reduced by using multiple channels. Consider the multichannel kernel-based time-encoding scheme shown in Figure~\ref{fig:schematicKernelBasedMultichannelSampling}, motivated by multichannel sampling schemes in \cite{chandra2008generalised, hormati2011common}. The input $x(t)$ is a $T$-periodic FRI signal (Eq.~\eqref{eq:signalModel}) and is sensed using $C$ time-encoding machines $\sT^{(1)}, \sT^{(2)}, \cdots, \sT^{(C)}$ and a kernel $g(t)$ that satisfies the alias-cancellation conditions (cf. Eq.~\eqref{eq:samplingKernel}). \\
\indent In the case of multichannel C-TEMs, let each channel have the reference $r^{(c)}(t) = A_{r}^{(c)} \cos(2\pi f_{r}^{(c)}t + \phi_{r}^{(c)})$, $c\in\llbracket 1, C \rrbracket$. The output of the $c^{\text{th}}$ channel is the sampling set $\sT_{\mathrm{CT}}^{(c)}y = \{t^{(c)}_{n} \; \vert \; y(t^{{(c)}}_{n}) = r^{(c)}(t^{(c)}_{n}), n\in\zz\}$ and samples of the signal $\{y(t^{(c)}_{n})\}_{n\in\zz}$ obtained using the references. Let $L^{(c)}$ be the number of samples measured by the $c^{\text{th}}$ channel in the interval $[0,T[$. As in the single-channel case (cf. Eq.~\eqref{eq:CTEMmeasurements}), the multichannel setup also gives rise to a linear system of equations $\by^{(c)} = \bG_{\mathrm{CT}}^{(c)} \hat{\bx}$ where $\by^{(c)} = [y(t^{(c)}_{1})  \cdots  y(t^{(c)}_{L^{(c)}})]^{\TT} \in \cc^{L^{(c)}}$ and $\bG_{\mathrm{CT}}^{(c)} \in\cc^{L^{(c)}\times N}$ for each channel. The definition of $\bG_{\mathrm{CT}}^{(c)}$ is similar to that given in Eq.~\eqref{eq:ctemMatrix}. \\
\indent Similarly, consider multichannel integrate-and-fire time-encoding, where each channel has parameters $b^{(c)}, \kappa^{(c)}$ and $\gamma^{(c)}$, $c\in\llbracket 1, C \rrbracket$ and possibly different initial values for the integrators. The output of the $c^{\text{th}}$ channel is the sampling set $\sT_{\mathrm{IF}}^{(c)}y = \{t^{(c)}_{n}\}_{n\in\zz}$ that satisfies Lemma~\ref{lem:ttransform} and Corollary~\ref{cor:iftemSamplingDensity} and local averages $\{\bar{y}^{(c)}_{n}\}_{n\in\zz}$ obtained using Lemma~\ref{lem:ttransform}. Let $L^{(c)}$ be the number of samples recorded by the $c^{\text{th}}$ channel. As in the single-channel case (cf. Eq.~\eqref{eq:iftemMeasurements}), the multichannel setup also gives rise to a linear system of equations $\bar{\by}^{(c)} = \bG_{\mathrm{IF}}^{(c)} \hat{\bx}$ where $\bar{\by}^{(c)} = [\bar{y}^{(c)}_{1}  \cdots  \bar{y}^{(c)}_{L^{(c)}}]^{\TT} \in \cc^{L^{(c)}}$ and $\bG_{\mathrm{IF}}^{(c)} \in\cc^{(L^{(c)}-1)\times N}$ for each channel. The definition of $\bG_{\mathrm{IF}}^{(c)}$ is similar to that given in Eq.~\eqref{eq:iftemMatrix}.\\
\indent In both multichannel flavours, we construct a concatenated system of linear equations as follows:
\begin{equation}
	\by \doteq \begin{bmatrix}
	\by^{(1)} \\ \by^{(2)} \\ \vdots \\ \by^{(C)}
	\end{bmatrix} =
	\begin{bmatrix}
	\bG^{(1)} \\ \bG^{(2)} \\ \vdots \\ \bG^{(C)}
	\end{bmatrix}
	\hat{\bx} \doteq \bG\hat{\bx}.
\label{eq:multichannelSystem}
\end{equation}
The system will admit a unique solution $\hat{\bx}$ when $\bG$ has full column-rank. This happens when the matrix is tall and when no two channels have identical trigger times. This can be achieved easily by setting different values for phase angle $\phi^{(c)} \in [0,2\pi[$ of the references in case of C-TEMs and setting different initial values for the integrators in case of IF-TEMs. Then, the sampling requirement in each channel can be reduced by the factor $C$.\\
\indent In the case of multichannel C-TEMs, the sampling requirement can be reduced by decreasing the frequency of the reference. The sufficient conditions for perfect reconstruction from multichannel C-TEM and multichannel IF-TEM follow.
\begin{proposition} (Sufficient condition for multichannel C-TEM)
Consider multichannel crossing-time-encoding of the $T$-periodic FRI signal $x(t)$ in Eq.~\eqref{eq:signalModel} with reference $r^{(c)}(t)$ and a sampling kernel $g(t)$ that satisfies Eq.~\eqref{eq:samplingKernel}. The sets of $C$ time instants $\{t^{(c)}_{n}\}_{n=1}^{L^{(c)}} \subset \sT^{(c)}_{\mathrm{CT}}(x*g) \cap [0,T[$, $c \in \llbracket 1, C \rrbracket$ is a sufficient representation of $x(t)$ when the reference of the $c^{\text{th}}$ channel chosen as $r^{(c)}(t) = A_{r}^{(c)} \cos(2\pi f_{r}^{(c)}t + \phi^{(r)})$ satisfies $A^{(c)}_{r} \geq \Vert x*g \Vert_{\infty}$ and $\displaystyle f^{(c)}_{r}\geq \frac{2K+1}{CT}, \; c \in \llbracket 1, C \rrbracket$.
\label{prop:ctemMultichannelGuarantees}
\end{proposition}
\begin{proposition} (Sufficient condition for multichannel IF-TEM)
Consider the $T$-periodic FRI signal $x(t)$ in Eq.~\eqref{eq:signalModel} encoded using multichannel IF-TEMs (cf. Figure~\ref{fig:schematicIFTEM}) with parameters $\{b^{(c)},\kappa^{(c)},\gamma^{(c)}\}$ and a sampling kernel $g(t)$ that satisfies Eq.~\ref{eq:samplingKernel}. The sets of $C$ time instants $\{t^{(c)}_{n}\}_{n=1}^{L^{(c)}} \subset \sT^{(c)}_{\mathrm{IF}}(x*g) \cap [0,T[$, $c \in \llbracket 1, C \rrbracket$ constitutes a sufficient representation of $x(t)$ with $(\sum_{c=1}^{C}L^{(c)}-C)\geq (2K+1)$ if the parameters of the $c^\text{th}$ channel satisfy:
\begin{equation*}
	\frac{\kappa^{(c)}\gamma^{(c)}}{b^{(c)}-\Vert x*g \Vert_{\infty}} < \frac{CT}{L^{(c)}}.
\end{equation*}
\label{prop:iftemMultichannelGuarantees}
\end{proposition}

%% -----------------------------------------------------------
%			RECOVERY IN PRESENCE OF NOISE
%% -----------------------------------------------------------

\section{Reconstruction in the Presence of Noise}
\label{sec:Noise_TEMFRI}
Depending on the circuit used in the time-encoding machine, in practice, one may have direct access to the sampling set as in the case of the method proposed by Naaman {\it et al.} \cite{naaman2021temhardware} or indirect measurements in the form of a bi-level signal whose transitions occur at the trigger times. The trigger times may be estimated from the bi-level signal using the sub-Nyquist method described in \cite{seelamantula2010sub} or the linear B-spline sampling method proposed in \cite{vetterli2002sampling}. Noise in the signal may perturb the trigger times. Even in the absence of noise, the trigger times can  be estimated only to a certain degree of accuracy. The jittered trigger times are expressed as
\begin{equation}
\tilde{t}_{n} = t_{n} + \nu_{n},
\label{eq:noiseModel}
\end{equation}
where the jitter $\nu_{n} \overset{\mathrm{i.i.d.}}{\sim} \mathcal{U}[-\sigma/2, \sigma/2]$ is i.i.d. uniformly distributed. The noise in the trigger times is different from the noise in the signal. Jitter in the temporal measurements permeates into the forward transformation matrix and the signal measurements.

%% -----------------------------------------------------------
\subsection{Effect of Noise in Crossing-Time-Encoding}
Consider kernel-based crossing-time-encoding of the FRI signal in Eq.~\eqref{eq:signalModel} with a sinusoidal reference $r(t) = A_{r}\cos(2\pi f_{r}t)$ that provides noisy trigger times
(Eq.~\eqref{eq:noiseModel}). The signal measurements are computed using the reference $y(\tilde{t}_{n}) = r(\tilde{t}_{n}) = r(t_{n}+\nu_{n})$. Since the reference is continuous and differentiable, the mean-value theorem suggests that there exists $t^{*}_{n}\in [t_{n}, t_{n}+\nu_{n}[$ such that $y(\tilde{t}_{n}) = r(t_{n}) + r'(t^{*}_{n})\nu_{n}$. Hence, the samples of the signal are random variables with
\begin{align}
\Ex[y(\tilde{t}_{n})] &= \Ex[r(t_{n}) + r'(t^{*}_{n})\nu_{n}] = r(t_{n}), \text{ and}\\
\varn(y(\tilde{t}_{n})) &= \varn(r(t_{n}) + r'(t^{*}_{n})\nu_{n}), \nonumber\\
&= \vert r'(t^{*}_{n}) \vert^{2} \frac{\sigma^{2}}{12} \leq (2\pi A_{r}f_{r})^{2}\frac{\sigma^{2}}{12}.
\end{align}
Since the variance of $y(\tilde{t}_n)$ depends on $t^*_n$, the distributions of $\{y(\tilde{t}_n)\}_{n\in\zz}$ are not identical, however, the expected value of $y(\tilde{t}_n)$ is equal to the true (noise-free) value $r(t_n)$.

%% -----------------------------------------------------------
\subsection{Effect of Noise in Integrate-and-Fire Time-Encoding}
Consider kernel-based time-encoding of the signal in Eq.~\eqref{eq:signalModel} using an IF-TEM with parameters $\{b, \kappa, \gamma\}$ (cf. Figure~\ref{fig:schematicIFTEM}) that provides noisy trigger times as in Eq.~\eqref{eq:noiseModel}. The local averages (using Lemma~\ref{lem:ttransform}) are given as $\tilde{\bar{y}}_{n} = -b(\tilde{t}_{n+1}-\tilde{t}_{n}) + \kappa\gamma$. The local averages are random variables with
\begin{align}
	\Ex[\tilde{\bar{y}}_{n}] &= \Ex[-b(\tilde{t}_{n+1}-\tilde{t}_{n}) + \kappa\gamma], \nonumber\\
	&= -b(t_{n+1} - t_{n}) + \kappa\gamma, \text{ and} \\
	\varn(\tilde{\bar{y}}_{n}) &= b^{2}(\varn(\nu_{n+1})+\varn(\nu_{n}))  = 2b^{2}\displaystyle\frac{\sigma^{2}}{12}.
\end{align}
The local averages are independent and identically distributed, with their expected value equal to the true values. The jitter in trigger times manifests as perturbations in the forward transformations $\bG_{\mathrm{CT}}$ and $\bG_{\mathrm{IF}}$, in  C-TEM and IF-TEM, respectively. The matrices may become ill-conditioned causing numerical instability in matrix inversion.
Naaman {\it et al.} \cite{naaman2021fritem} consider IF-TEM and
suggest removing the $(M+1)^\text{th}$ column in $\bG_{\mathrm{IF}}$ to improve the conditioning of the forward transformation. Their technique recovers $\{\hat{x}_{m}|m\in\llbracket -M, M \rrbracket\setminus \{0\}\}$, i.e., the zeroth Fourier coefficient cannot be recovered. They show, by setting $\hat{x}_{0} = 0$, that Prony's method produces accurate estimates with oversampling.\\
\indent In this paper, we treat reconstruction in the presence of measurement noise as solving for the Fourier coefficients that satisfy a linear system of equations of the type $\by = \bG\hat{\bx}$, where $\bG$ is the forward transformation and $\by$ contains measurements of $\sE y$, in addition to the constraint that the annihilating filter $\bh\in\cc^{K+1}$ lies in the null space of $\Gamma_{K}\hat{\bx}$. As discussed in Section~\ref{subsec:pronysmethod}, the roots of the annihilating filter are in one-to-one correspondence with the shifts in Eq.~\eqref{eq:signalModel}. This system is representative of the one encountered in the case of C-TEM (Section~\ref{sec:CTEM_sampling}) or IF-TEM (Section~\ref{sec:IFTEM_sampling}) or multichannel TEM (Section~\ref{sec:MTEMFRI}). In the presence of noise, the matrix $\Gamma_{K}\hat{\bx}$ becomes full rank, whereas the noise-free counterpart is rank-deficient. In standard FRI sampling, the matrix $\Gamma_{K}\hat{\bx}$ is subjected to a rank-restoration process by denoising the measurements using Cadzow's technique \cite{cadzow}. The technique falls in the broad category of structured low-rank approximation (SLRA) methods as pointed out in \cite{condat2015cadzow}. The rank-restoration strategies are decoupled from the procedure to obtain the Fourier coefficients. Pan {\it et al.} \cite{pan2016towards} considered a generalized denoising problem and solved jointly for the Fourier coefficients and the annihilating filter by incorporating the annihilation property as a constraint in the optimization --- this is referred to as the ``Generalized FRI'' approach. We adopt a similar approach for the time-encoding problem at hand. More precisely, we solve the system in $\by = \bG\hat{\bx}$ considering noisy time-encoded measurements, together with $(\Gamma_{K}\hat{\bx})\bh = \zerovec$. In the case of C-TEM, the problem  reduces to the form considered by Pan {\it et al.}, whereas in the case of IF-TEM, the signal measurements are obtained after integration. \\
\indent To avoid the trivial solution $\bh=\zerovec$ for the annihilating filter, suitable regularization becomes necessary. Do\v{g}an {\it et al.} \cite{dougan2015reconstruction} presented empirical evidence to argue that the constraint $\langle \bh, \bh^{(0)} \rangle = 1$, with $\bh^{(0)} \in \cc\mathcal{N}(\zerovec, \mathbf{I})$ being the initialization for the iterative algorithm, works better than other candidate choices such as $h_{0} = 1$ or $\Vert \bh \Vert_{2} = 1$. The corresponding optimization program takes the form
\leqnomode
\begin{equation}
\begin{split}
	\underset{\hat{\bx}\in\cc^N,\, \bh\in\cc^{K+1}}{\text{minimize }}&  \Vert \bG\hat{\bx}-\by \Vert_2^2 \\
	\text{subject to }& (\Gamma_K\hat{\bx})\bh = \zerovec, \\
	& \langle \bh, \bh^{(0)} \rangle = 1.
\end{split}
\tag{\textbf{P}}\label{eq:genfri}
\end{equation}
The program is nonconvex when the optimization is carried jointly over $\hat{\bx}$ and $\bh$. However, holding $\bh$ fixed makes the problem convex in $\hat{\bx}$ and vice versa, which suggests that an {\it alternating minimization} strategy could be deployed.\\
\indent Consider the subprogram where $\bh$ is held fixed in \eqref{eq:genfri}. The annihilation constraint can be expressed in the variable $\hat{\bx}$ using the right dual of $\Gamma_{K}$ (from the discussion following Definition~\ref{def:toep} in Section~\ref{sec:preliminaries}) resulting in the optimization:
\begin{equation}
\begin{split}
	\underset{\hat{\bx}\in\cc^N}{\text{minimize }}& \Vert \bG\hat{\bx}-\by \Vert_2^2, \\
	\text{subject to }& (\zeta_{N}\bh) \hat{\bx} = \zerovec,
\end{split}
\tag{\textbf{P1}}\label{eq:xGenfri}
\end{equation}
which is a quadratic program with a linear equality constraint. The minimizer of \eqref{eq:xGenfri} is given in closed-form as (cf. Chapter 5, \cite{boyd2004optimisation}):
\begin{equation}
	\hat{\bx} = \bbeta - (\bG^{\HH}\bG)^{-1}(\zeta_{N}\bh)^{\HH} \bPsi^{-1} (\zeta_{N}\bh)\bbeta,
\label{eq:xMinimiser}
\end{equation}
where $\quad \bPsi = \left[ (\zeta_{N}\bh)(\bG^{\HH}\bG)^{-1} (\zeta_{N}\bh)^{\HH}\right]$ and $\bbeta = (\bG^{\HH}\bG)^{-1}\bG^{\HH}\by$.
The minimum value of the objective function in \eqref{eq:xGenfri} can be shown to be equal to $\bh^{\HH}(\Gamma_{K}\bbeta)^{\HH}\bPsi^{-1}(\Gamma_{K}\bbeta)\bh + \text{terms independent of } \bh$ (cf. Appendix A, \cite{pan2016towards}). Substituting the optimum $\hat{\bx}$ from Eq.~\eqref{eq:xMinimiser} in \eqref{eq:genfri} gives rise to a quadratic subprogram in $\bh$ with an affine equality constraint:
\begin{equation}
\begin{split}
	\underset{\bh\in\cc^{(K+1)}}{\text{minimize }}& \bh^{\HH}(\Gamma_{K}\bbeta)^{\HH}\bPsi^{-1}(\Gamma_{K}\bbeta)\bh,\\
	\text{subject to }& \langle \bh, \bh^{(0)} \rangle = 1.
\end{split}
\tag{\textbf{P2}}\label{eq:hGenfri}
\end{equation}
\reqnomode
The matrix $\bPsi$ is considered independent of $\bh$ with the evaluation of the matrix from the update in Eq.~\eqref{eq:xMinimiser}. The minimizer of \eqref{eq:hGenfri} has a closed-form expression given by (cf. Chapter 5, \cite{boyd2004optimisation})
\begin{align}
	\bh = \frac{\left[ (\Gamma_{K}\bbeta)^{\HH}\bPsi^{-1}(\Gamma_{K}\bbeta) \right]^{-1}}{(\bh^{(0)})^{\HH}\left[(\Gamma_{K}\bbeta)^{\HH}\bPsi^{-1}(\Gamma_{K}\bbeta) \right]^{-1}\bh^{(0)}} \bh^{(0)}.
\end{align}
\indent The minimizer of \eqref{eq:genfri} can be found using alternating minimization of \eqref{eq:xGenfri} and \eqref{eq:hGenfri}. The attractive feature is that each step has a closed-form solution. The downside is that the solutions involve matrix inversions, which are not only computationally demanding, but also prone to numerical instability \cite{pan2016towards}. Pan {\it et al.} suggested that the minimizers in each subprogram can be expressed as solutions to larger linear systems of equations using slack variables, which is presented next.
\begin{proposition} The optimization in \eqref{eq:xGenfri} can be carried out by solving the following linear system of equations:
\begin{equation}
	\begin{bmatrix}
		\bG^{\HH}\bG & (\zeta_{N}\bh)^{\HH} \\
		\zeta_{N}\bh & \zerovec \\
	\end{bmatrix}
	\begin{bmatrix}
		\hat{\bx} \\ \boldsymbol{\alpha}
	\end{bmatrix} =
	\begin{bmatrix}
		\bG^{\HH}\by \\ \zerovec
	\end{bmatrix}.
\label{eq:xUpdate}
\end{equation}
Similarly, the optimization in \eqref{eq:hGenfri} can be carried out by solving the following linear system of equations
\begin{equation}
	\begin{bmatrix}
		\zerovec & (\Gamma_{K}\bbeta)^{\HH} & \zerovec & \bh^{(0)} \\
		\Gamma_{K}\bbeta & \zerovec & -\zeta_{N}\bh & \zerovec \\
		\zerovec & -(\zeta_{N}\bh)^{\HH} & \bG^{\HH}\bG & \zerovec \\
		(\bh^{(0)})^{\HH} & \zerovec & \zerovec & 0
	\end{bmatrix}
	\begin{bmatrix}
		\bh \\ \boldsymbol{\alpha} \\ \boldsymbol{\xi} \\ \lambda
	\end{bmatrix} =
	\begin{bmatrix}
		\zerovec \\ \zerovec \\ \zerovec \\ 1
	\end{bmatrix},
\label{eq:hUpdate}
\end{equation}
where $\boldsymbol{\alpha}, \boldsymbol{\xi}$, and $\lambda$ are the slack variables.
\end{proposition}
The proof is provided in \cite{pan2016towards}. The solution to program~\eqref{eq:genfri} is obtained by alternating between the programs \eqref{eq:xGenfri} and \eqref{eq:hGenfri}. Although there are no provable convergence guarantees for the alternating minimization strategy employed here, the reformulation gives rise to a stable approach and techniques such as Gaussian elimination can be used. \\
\indent A suitable stopping criterion must be employed to terminate the iterations. Typically, the iterations are terminated when the value of the objective function $\Vert \by - \bG\hat{\bx} \Vert^{2}_{2}$ falls below a certain threshold or when the maximum iteration count is reached. Since the overall program~\eqref{eq:genfri} is nonconvex, it is possible that the stopping criteria may not be met for certain choices of $\bh^{(0)}$. In such cases, the algorithm is restarted with a different initialization. We found that, with  {\it multiple restarts}, the technique converged in about $50$ iterations. Let $(\hat{\bx}^{\mathrm{opt}},\bh^{\mathrm{opt}})$ be the solution obtained after convergence. The shifts $\{\tau_{k}\}_{k=0}^{K-1}$ are computed from the roots of $\bh^{\mathrm{opt}}$ and the amplitudes $\{c_{k}\}_{k=0}^{K-1}$ are computed using linear regression using $\hat{\bx}^{\mathrm{opt}}$ and the shifts obtained (cf. Eq.~\eqref{eq:swce_pulse}). We refer to this method, which is motivated by the Generalized FRI solver of Pan {\it et al.} \cite{pan2016towards}, as ``GenFRI-TEM.'' The steps are listed in  Algorithm~\ref{algo:genFRI}.
\begin{algorithm}[t]
	\KwIn{Time-encoding $\{t_{n}\}_{n=1}^{L} \subset \sT (x*g)$, samples $(\sE (x*g))(\{t_{n}\}_{n=1}^{L})$, stopping criterion $\eta$, max. iterations $N^{\#}$, and number of restarts}
	\KwOut{Amplitudes $\{c_{k}\}_{k=0}^{K-1}$ and shifts $\{\tau_{k}\}_{k=0}^{K-1}$}
	\Repeat{number of restarts}{
	Set $n = 0$ \;
	Initialize $\bh^{(0)} \in \cc^{(K+1)}$ randomly \;
  	\Repeat{$\Vert \by - \bG\hat{\bx} \Vert_{2}^{2} \leq \eta$ or $n > N^{\#}$}{
  	Update $\hat{\bx}^{(n)}$ by solving Eq.~\eqref{eq:xUpdate} \;
    Update $\bh^{(n)}$ by solving Eq.~\eqref{eq:hUpdate} \;
		$n \leftarrow n + 1$ \;
    	}}
	Compute $\{\tau_{k}\}_{k=0}^{K-1}$ from the roots of $\mathbf{h}^{(n)}$\;
	Compute $\{c_{k}\}_{k=0}^{K-1}$ by linear regression.
\caption{GenFRI-TEM: Generalized FRI for signal reconstruction from time-encoded measurements.}
\label{algo:genFRI}
\end{algorithm}
%% -----------------------------------------------------------
%					RESULTS
%% -----------------------------------------------------------

\section{Simulation Results}
\label{sec:experiments}
We validate GenFRI-TEM through simulations, under noise-free and noisy, single-channel as well as multichannel measurements and compare the performance with the benchmark reconstruction methods. In keeping with the spirit of reproducible research \cite{vetterli2009reproducible}, we have made our codes available on GitHub \cite{kamath2021genfritem}.\\
\indent Consider a $T$-periodic FRI signal $x(t)$ with period $T=1$ comprising Dirac impulses:
\begin{align}
	x(t) &= \sum_{m\in\zz}\sum_{k=0}^{4} c_{k} \delta(t-\tau_{k}-m),
	% x_{2}(t) &= \sum_{m\in\zz}\sum_{k=0}^{2} c_{k} \varphi(t-\tau_{k}-m),
\label{eq:expDiracSignal}
\end{align}
where the amplitudes are drawn from the Gaussian distribution $\mathcal{N}(0.5,1)$ and the shifts are selected uniformly at random over $[0,1[$. Once selected, the parameters are kept fixed in all the realizations.
% The pulse in $x_2$ is chosen as $\varphi(t) = \beta^{(3)}(10t)$, where $\beta^{(3)}$ is the cubic B-spline. The amplitudes are $\{c_{k}\}_{k=0}^{2} = \{0.5, -0.35, 0.3\}$ and the shifts are $\{\tau_{k}\}_{k=0}^{2} = \{0.20, 0.34, 0.74\}$.
% with $\{c_k\}_{k=0}^4 = \{0.25535916, 0.79226309, -0.27083644,  0.99033927, -0.53013379\}$ and $\{\tau_k\}_{k=0}^4 = \{0.13011428, 0.46690178, 0.53536679, 0.81987613, 0.89174464\}$.
The sampling kernel (cf. Figure~\ref{fig:schematicKernelBasedSampling}) is chosen as
\begin{equation*}
g(t) = \sum_{k=-K}^{K}e^{\jj \omega_0 k t} = \displaystyle \frac{\sin((K+0.5)\omega_{0} t)}{\sin(\omega_{0}t/2)}, -\frac{T}{2}\leq t \leq \frac{T}{2},
\end{equation*}
which is the sum-of-sincs (SoS) kernel in the Fourier domain \cite{tur2011innovation}. This choice considers $M=K$ in Eq.~\eqref{eq:filteredSignal}, which gives access to $2K+1$ Fourier coefficients via time-domain measurements. In each case, we use Algorithm~\ref{algo:genFRI} with a maximum of $50$ iterations and allowing up to $50$ restarts.
%% -----------------------------------------------------------
\subsection{Reconstruction from Noise-Free Measurements}
\label{subsec:single_channel_noise_free_experiments}
\begin{figure}[t]
\centering
\subfigure[]{\label{fig:ctem_clean_a}\includegraphics[width=3.35in]{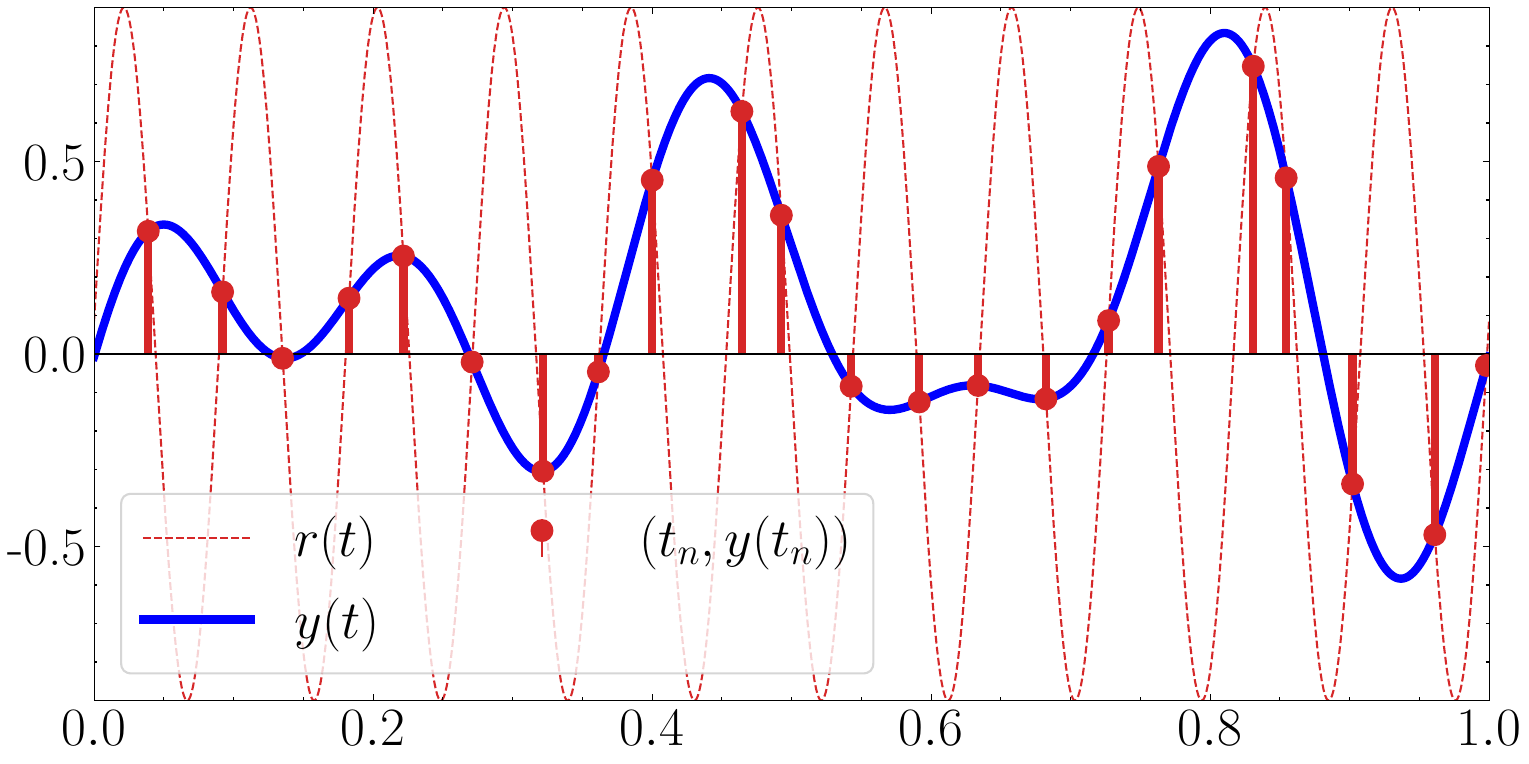}}
\subfigure[]{\label{fig:ctem_clean_b}\includegraphics[width=3.35in]{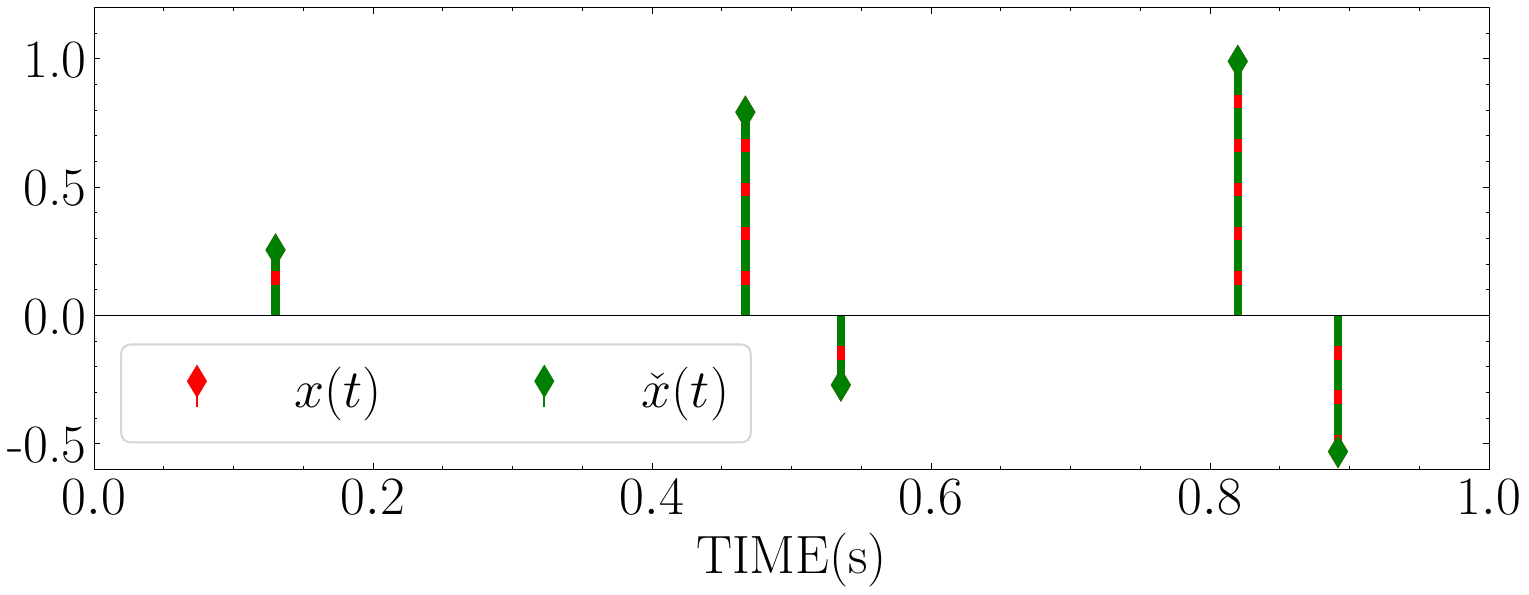}}
\caption{Time-encoding and reconstruction using C-TEM. (a) shows the filtered signal $y(t)$, reference signal $r(t)$, trigger times $\{t_n\}$; and (b) shows the input $x(t)$ and its reconstruction $\check{x}(t)$.}
\label{fig:ctem_clean}
\end{figure}

\begin{figure}[t]
\centering
\subfigure[]{\label{fig:iftem_clean_a}\includegraphics[width=3.35in]{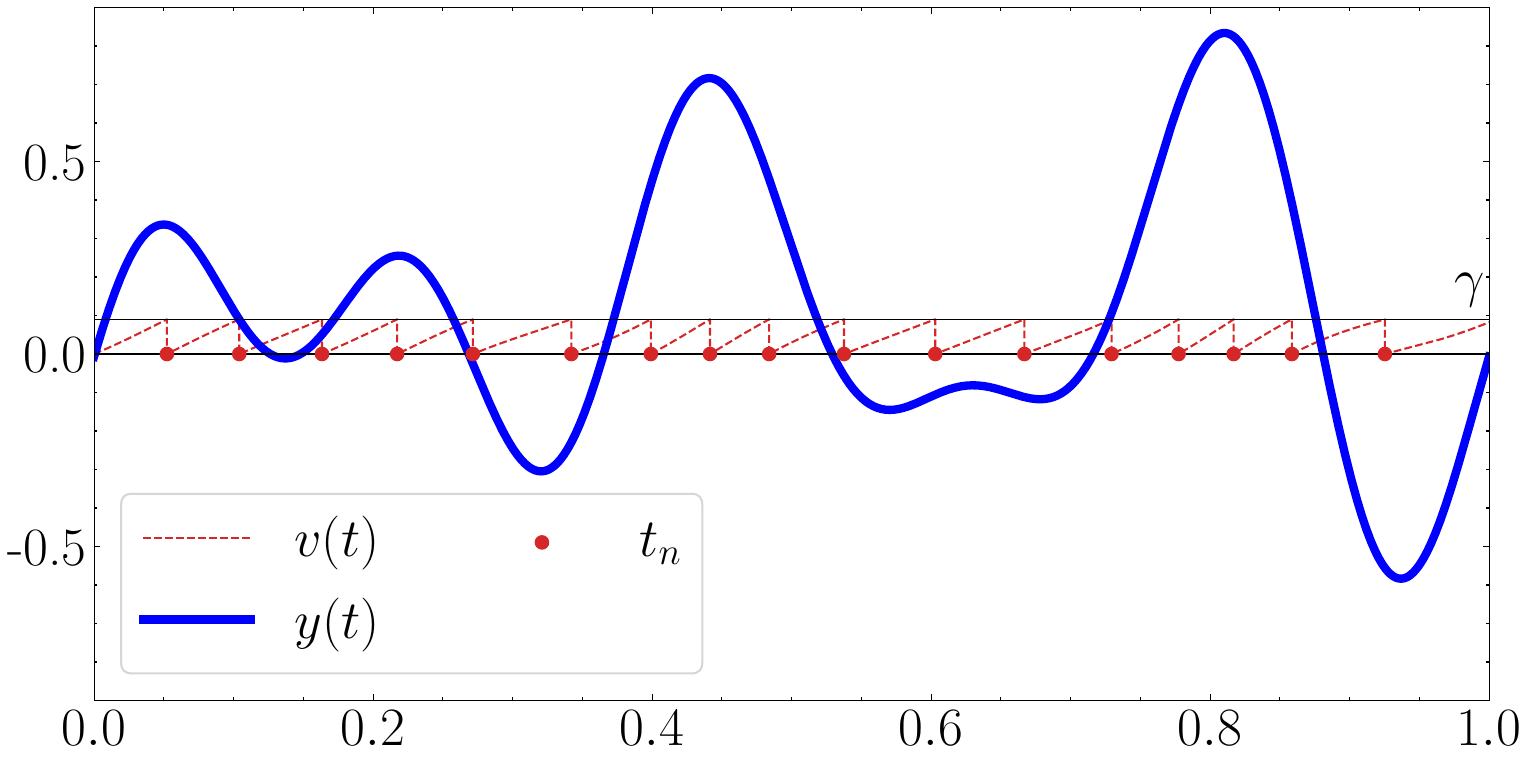}}
\subfigure[]{\label{fig:iftem_clean_b}\includegraphics[width=3.35in]{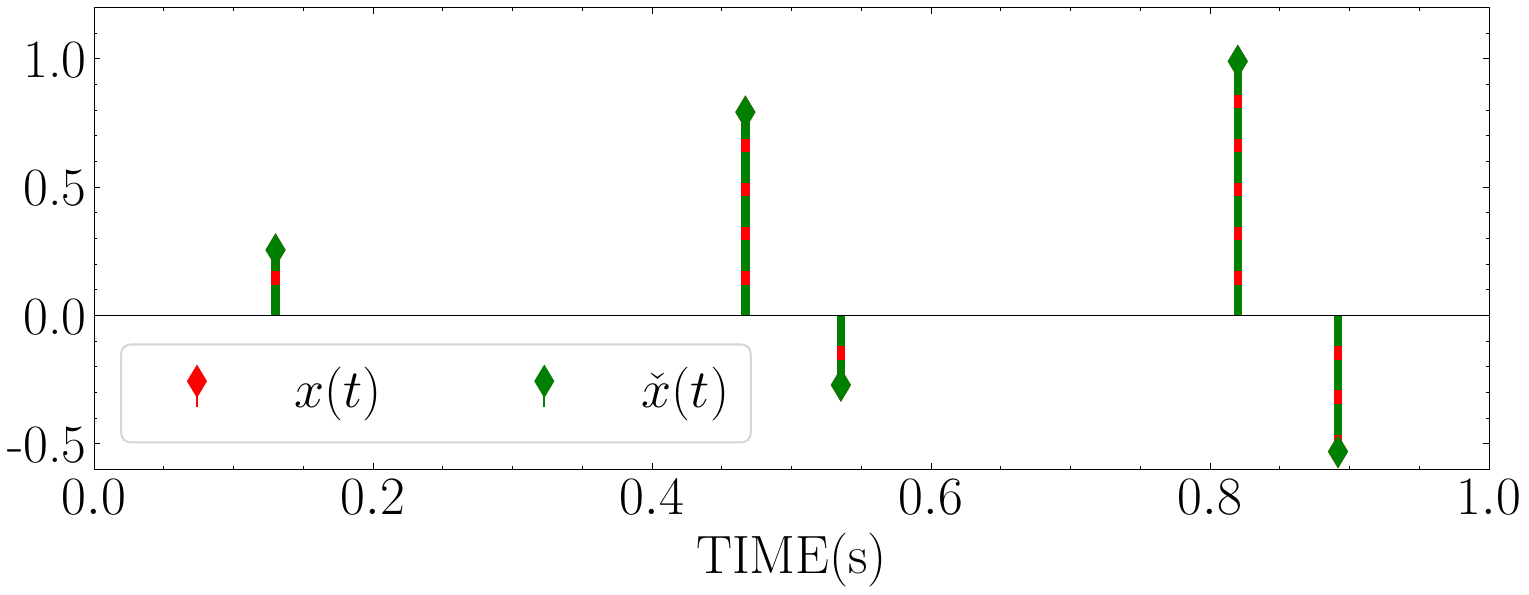}}
\caption{Time-encoding and reconstruction using IF-TEM: (a) shows the filtered signal $y(t)$, the output of the integrator $v(t)$, trigger times $\{t_n\}$; and (b) shows the input $x(t)$ and its reconstruction $\check{x}(t)$.}
\label{fig:iftem_clean}
\end{figure}

Consider the single-channel case and time-encoded measurements of $x(t)$ using C-TEM with reference $r(t) = 0.9 \cos(2\pi\cdot 11 \cdot t + \phi_r)$, and IF-TEM with parameters $\{b,\kappa,\gamma\} = \{1.5, 1, 0.09\}$. The phase angles $\phi_r$ for the reference signals in C-TEM are selected uniformly at random over $[0,2\pi[$.
%\begin{table}[t]
%\centering
%\caption{Parameters of single-channel TEMs used in the simulations.}
%\begin{tabular}{ c | c | ccc }
%    \toprule
%    \multirow{2}{3em}{Signal} & C-TEM & \multicolumn{3}{c}{IF-TEM} \\
%    & $r(t)$ & $b$ & $\kappa$ & $\gamma$ \\
%    \midrule
%    $x$ & $0.9 \cos(2\pi\cdot 11 \cdot t + \phi_r)$ & $1.5$ & $1$ & $0.09$ \\
%    % \midrule
%    % $x_{2}$ & $0.3 \cos(2\pi\cdot 7 \cdot t + \phi_r)$ & $1.3$ & $1$ & $0.09$ \\
%    \bottomrule
%\end{tabular}
%\label{tab:singleChannel}
%\end{table}
The parameters are chosen such that they critically satisfy the sampling requirement (cf. Propositions~\ref{prop:ctemGuarantees} and ~\ref{prop:iftemGuarantees}). Figure~\ref{fig:ctem_clean} shows C-TEM based reconstruction of $x(t)$. Likewise,
Figure~\ref{fig:iftem_clean} shows IF-TEM based reconstruction of $x(t)$. The reconstruction is accurate up to numerical precision in both encoding schemes.

%\begin{table}[t]
%\centering
%\caption{Parameters of multichannel TEMs used in the simulations.}
%\begin{tabular}{ c | c | ccc }
%    \toprule
%    \multirow{2}{3em}{Signal} & C-TEM & \multicolumn{3}{c}{IF-TEM} \\
%    & $r^{(c)}(t)$ & $b^{(c)}$ & $\kappa^{(c)}$ & $\gamma^{(c)}$ \\
%    \midrule
%    \multirow{2}{3em}{$x$} & $0.9 \cos(2\pi\cdot 5.5 \cdot t + \phi^{(1)}_{r})$ & $1.5$ & $1$ & $0.18$ \\
%    & $0.9 \cos(2\pi\cdot 5.5 \cdot t + \phi^{(2)}_{r})$ & $1.5$ & $0.82$ & $0.22$ \\
%    % \midrule
%    % \multirow{2}{3em}{$x_{2}$} & $0.4 \cos(2\pi\cdot 3.5 \cdot t + \phi^{(1)}_{r})$ & $1.3$ & $1$ & $0.18$ \\
%    % & $0.4 \cos(2\pi\cdot 3.5 \cdot t + \phi^{(2)}_{r})$ & $1.3$ & $0.82$ & $0.22$ \\
%    \bottomrule
%\end{tabular}
%\label{tab:multiChannel}
%\end{table}

\begin{figure}[!t]
\centering
\subfigure[]{\label{fig:ctem2_clean_a}\includegraphics[width=3.35in]{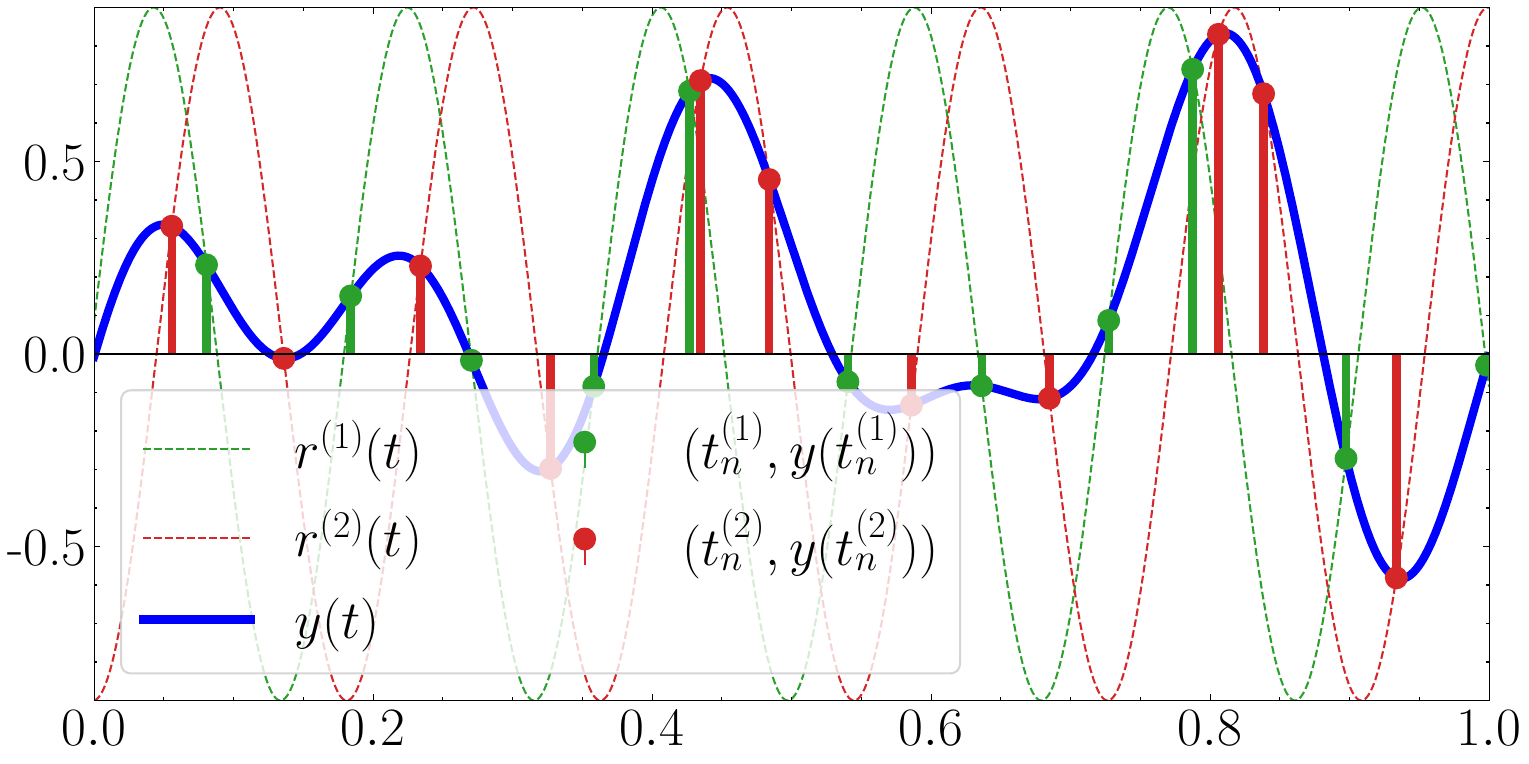}}
\subfigure[]{\label{fig:ctem2_clean_b}\includegraphics[width=3.35in]{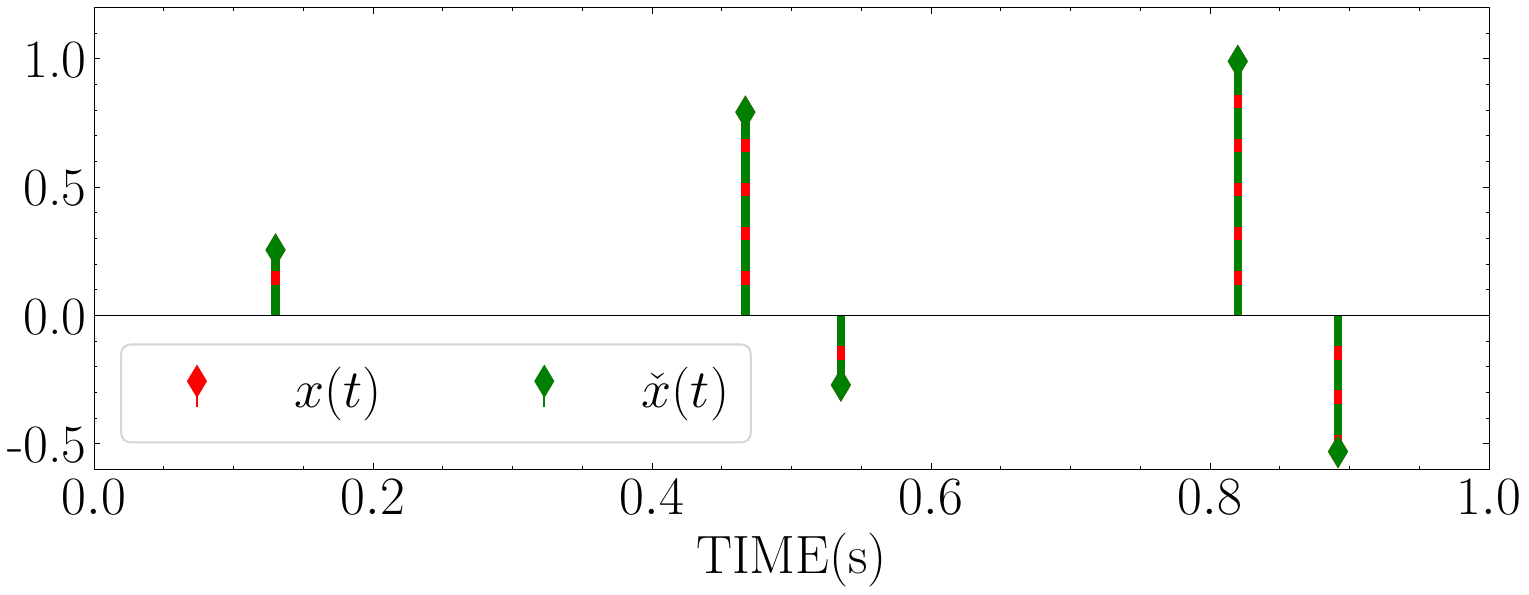}}
\caption{Time-encoding and reconstruction using C-TEM: (a) shows the filtered signal $y(t)$, the reference signals $r^{(1)}(t)$, $r^{(2)}(t)$ and the corresponding trigger times; and (b) shows the input stream of Dirac impulses $x(t)$ and reconstruction $\check{x}(t)$.}
\label{fig:ctem2_clean}
\end{figure}

\begin{figure}[!t]
\centering
\subfigure[]{\label{fig:iftem2_clean_a}\includegraphics[width=3.35in]{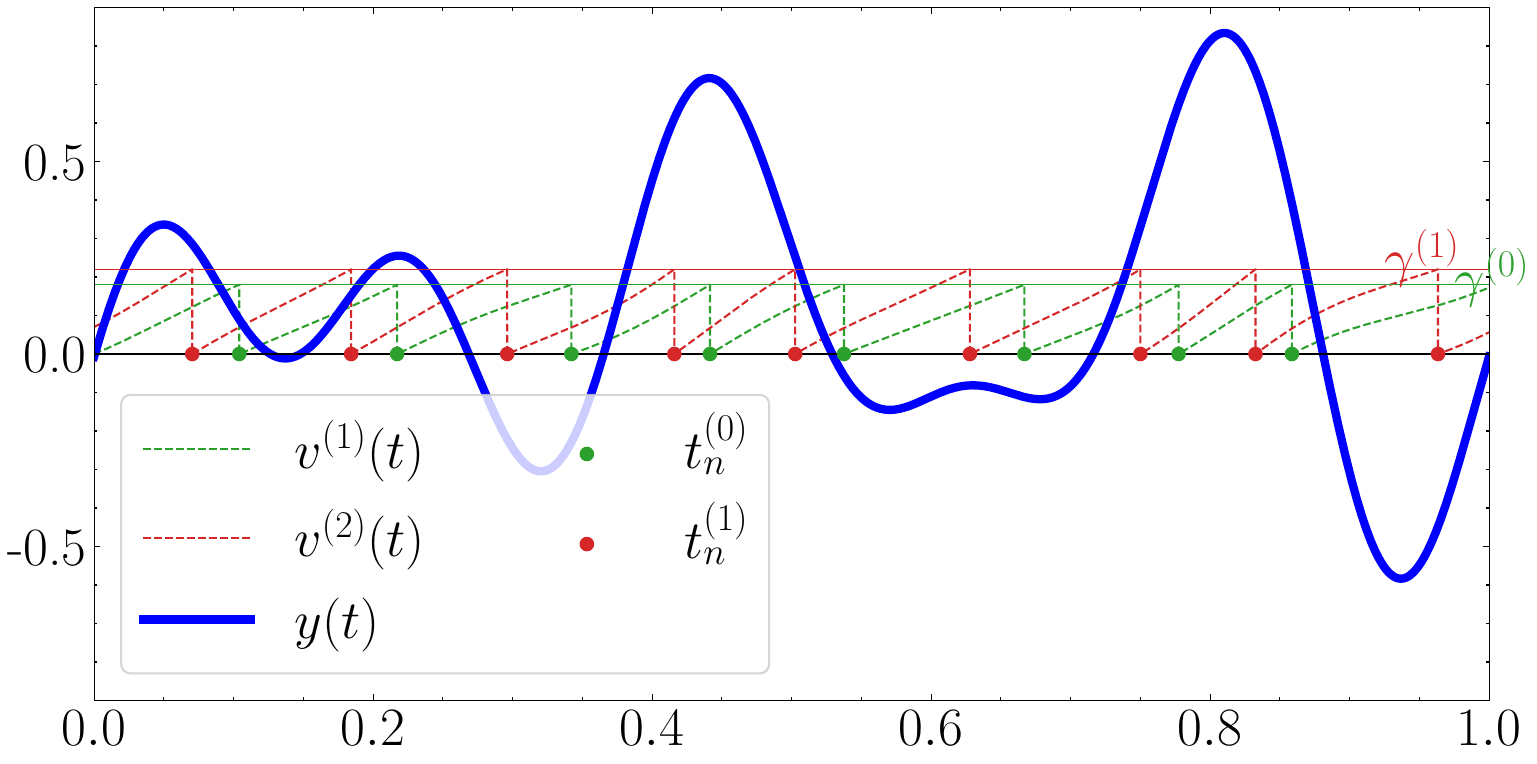}}
\subfigure[]{\label{fig:iftem2_clean_b}\includegraphics[width=3.35in]{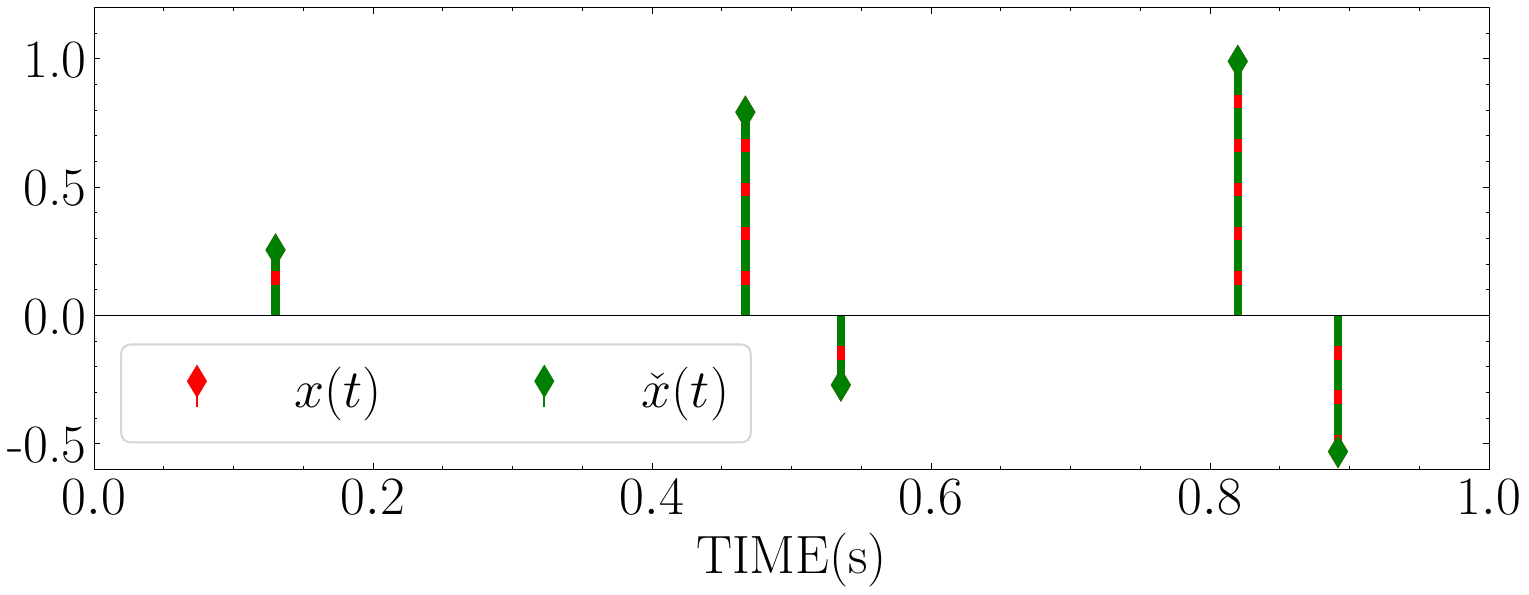}}
\caption{Time-encoding and reconstruction using IF-TEM: (a) shows the filtered signal $y(t)$, the output of the integrators $v^{(1)}(t)$, $v^{(2)}(t)$ and the corresponding trigger times; and (b) shows the input stream of Dirac impulses $x(t)$ and its reconstruction $\check{x}(t)$.}
\label{fig:iftem2_clean}
\end{figure}
Next, we consider two-channel encoding using C-TEMs with references $r^{c}(t) = 0.9 \cos(2\pi\cdot 5.5 \cdot t + \phi^{(c)}_{r}), \; c=1,2$ and IF-TEMs with parameters $\{b^{(1)},\kappa^{(1)},\gamma^{(1)}\} = \{1.5, 1, 0.18\}$, and $\{b^{(2)},\kappa^{(2)},\gamma^{(2)}\} = \{1.5, 0.82, 0.22\}$. The phase angles $\phi^{(1)}_{r}$ and $\phi^{(2)}_{r}$ for the reference signals in C-TEM are independently selected uniformly at random over $[0,2\pi[$. The initial values of the integrators in the two channels of IF-TEM are set to $0$ and $0.07$. This ensures that the two channels have distinct sampling sets. The parameters are chosen such that they critically satisfy the sampling requirements (cf. Propositions~\ref{prop:ctemMultichannelGuarantees} and ~\ref{prop:iftemMultichannelGuarantees}). Note the reduction in sampling requirement in each channel as opposed to the single-channel case. Figure~\ref{fig:ctem2_clean} shows multichannel C-TEM based reconstruction of $x(t)$. Likewise, Figure~\ref{fig:iftem2_clean} shows multichannel IF-TEM based reconstruction of $x(t)$. The reconstruction is accurate up to numerical precision in both encoding schemes.

%% -----------------------------------------------------------

\begin{figure}[t]
\centering
\subfigure[]{\label{fig:ctem_noise_rec}\includegraphics[width=3.35in]{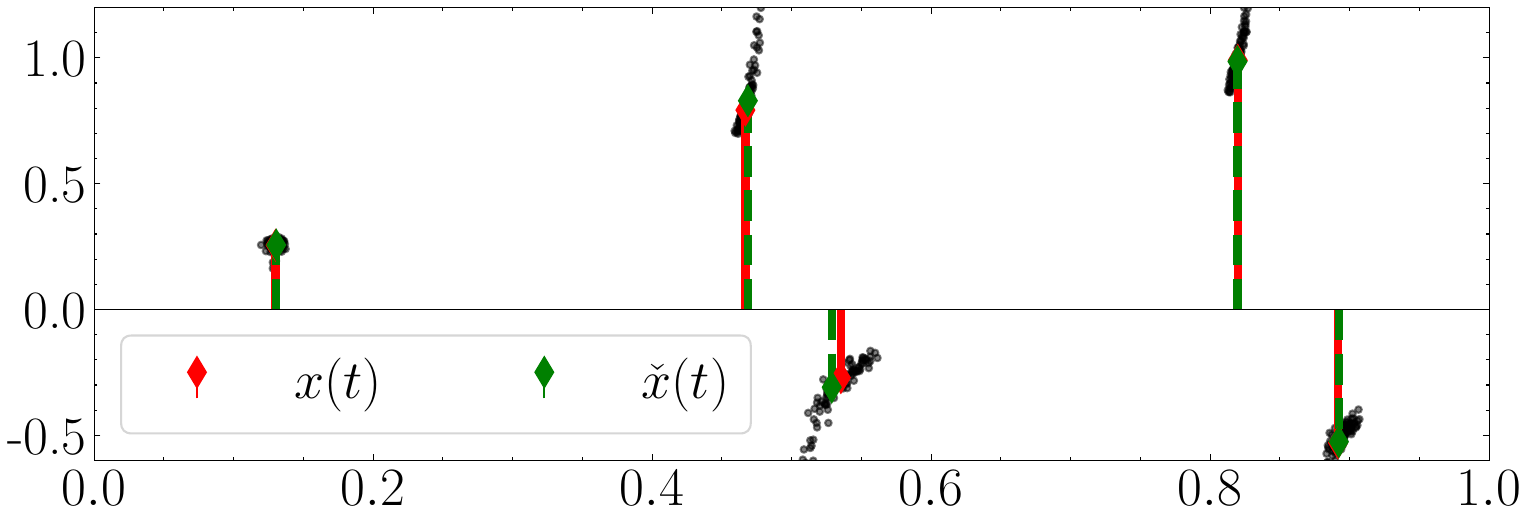}}
\subfigure[]{\label{fig:iftem_noise_rec}\includegraphics[width=3.35in]{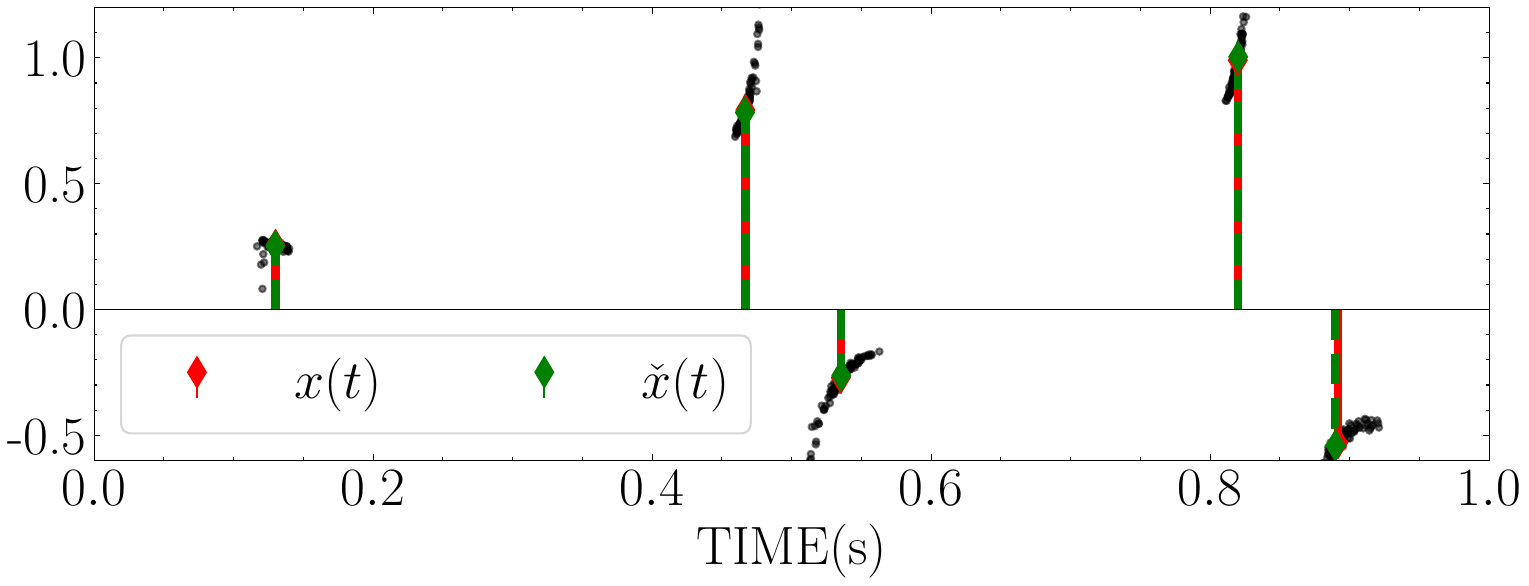}}
\caption{Time-encoding of an input stream of Dirac impulses $x(t)$ using (a) C-TEM, and (b) IF-TEM and the median reconstruction $\check{x}(t)$ in the presence of jitter with $\sigma=10^{-3}$. The scatter plot shows the reconstruction for $100$ realizations of jitter.}
\label{fig:rec_noisy}
\end{figure}
\begin{figure}[t]
\centering
\subfigure[]{\label{fig:ctem_shift_nmse}\includegraphics[width=3.5in]{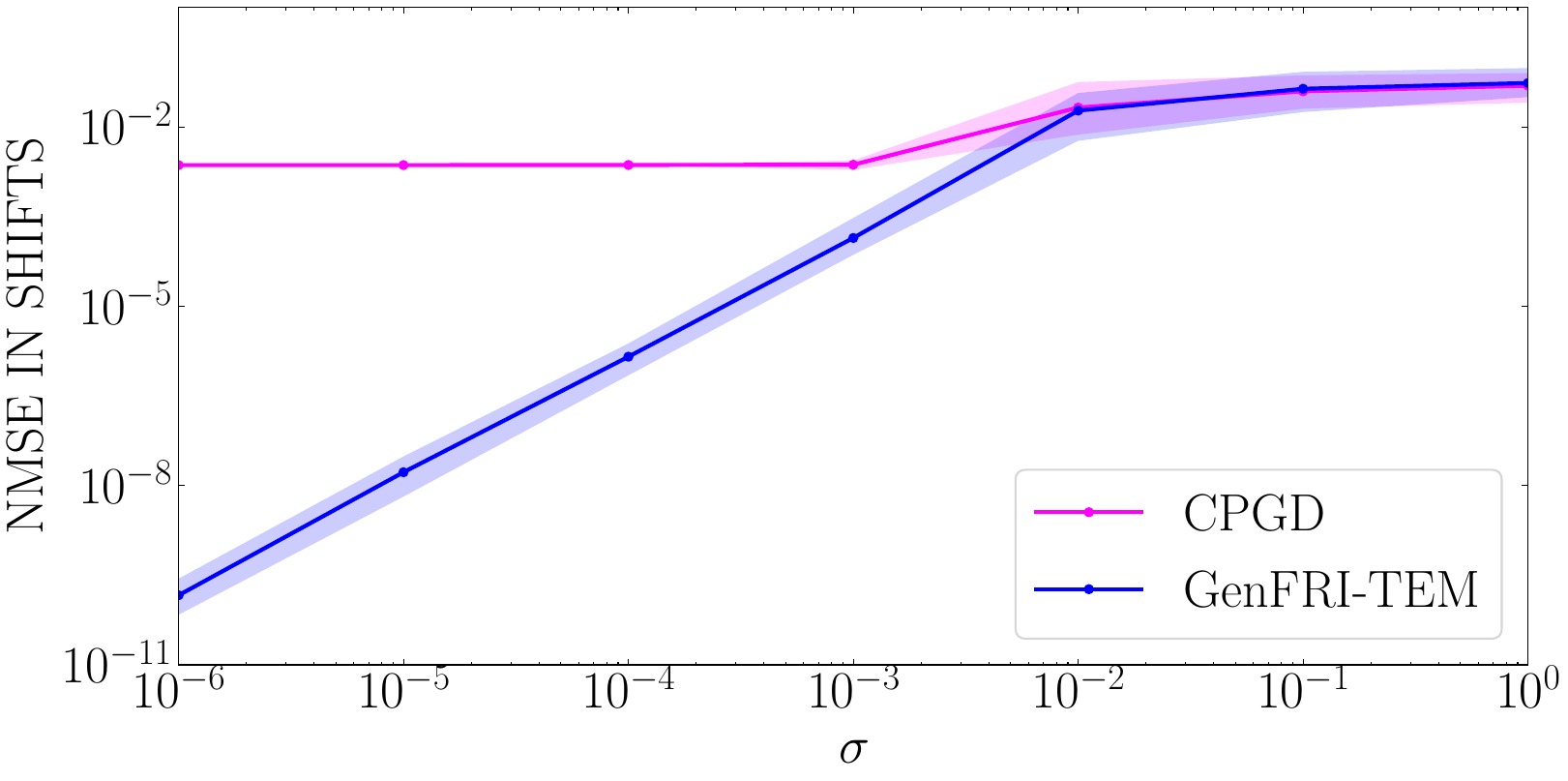}}
\subfigure[]{\label{fig:ctem_amp_nmse}\includegraphics[width=3.5in]{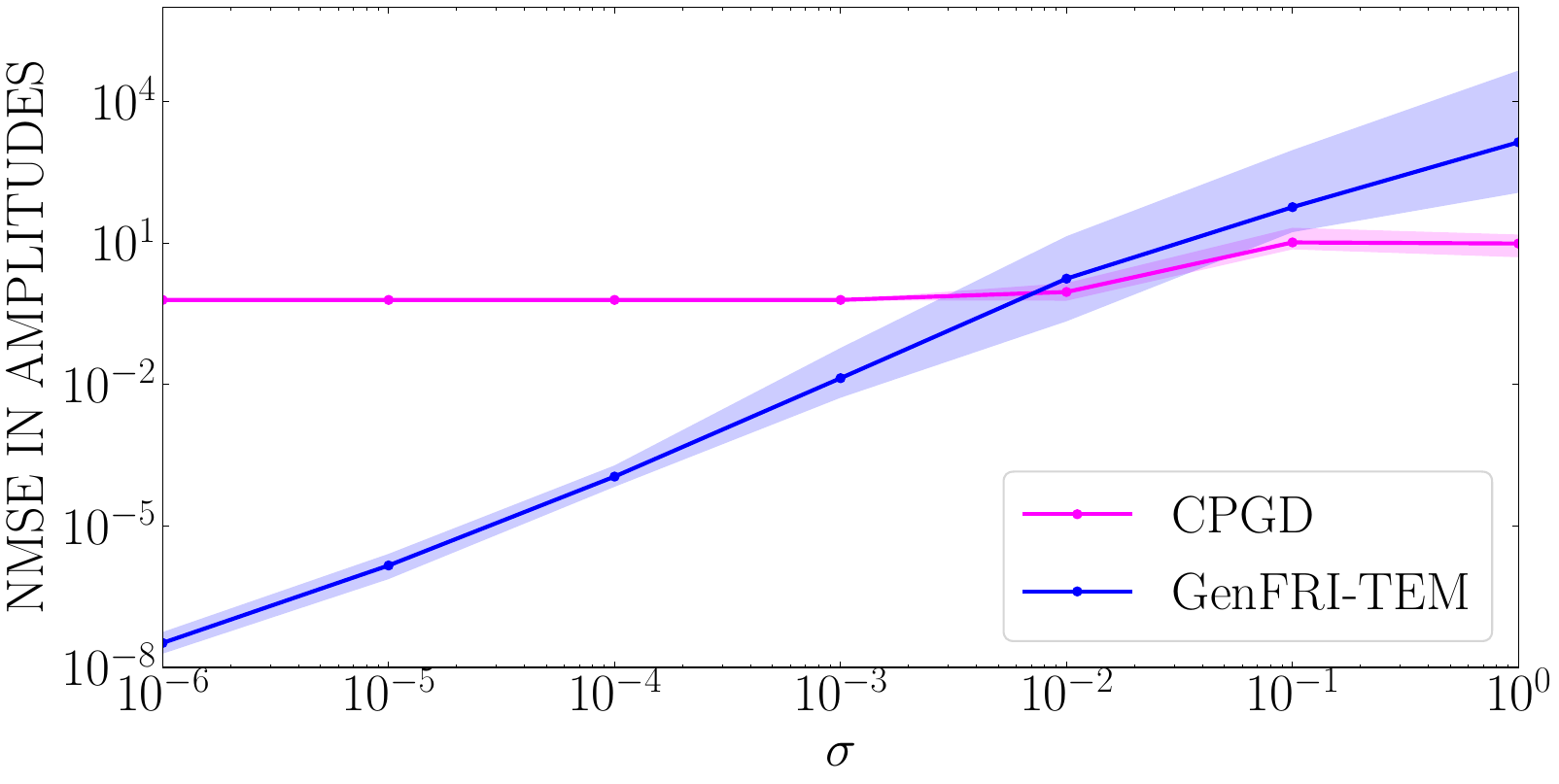}}
\caption{Normalized mean-square error in estimating the amplitude and shift parameters as a function of the noise level $\sigma$ from C-TEM measurements. The solid line shows the median error and the shaded region captures the inter-quartile region of the estimate. The results are over $100$ realizations of the jitter. The axes are in log scale.}
\label{fig:ctem_errors_noisy}
\end{figure}
\begin{figure}[t]
\centering
\subfigure[]{\label{fig:iftem_shift_nmse}\includegraphics[width=3.5in]{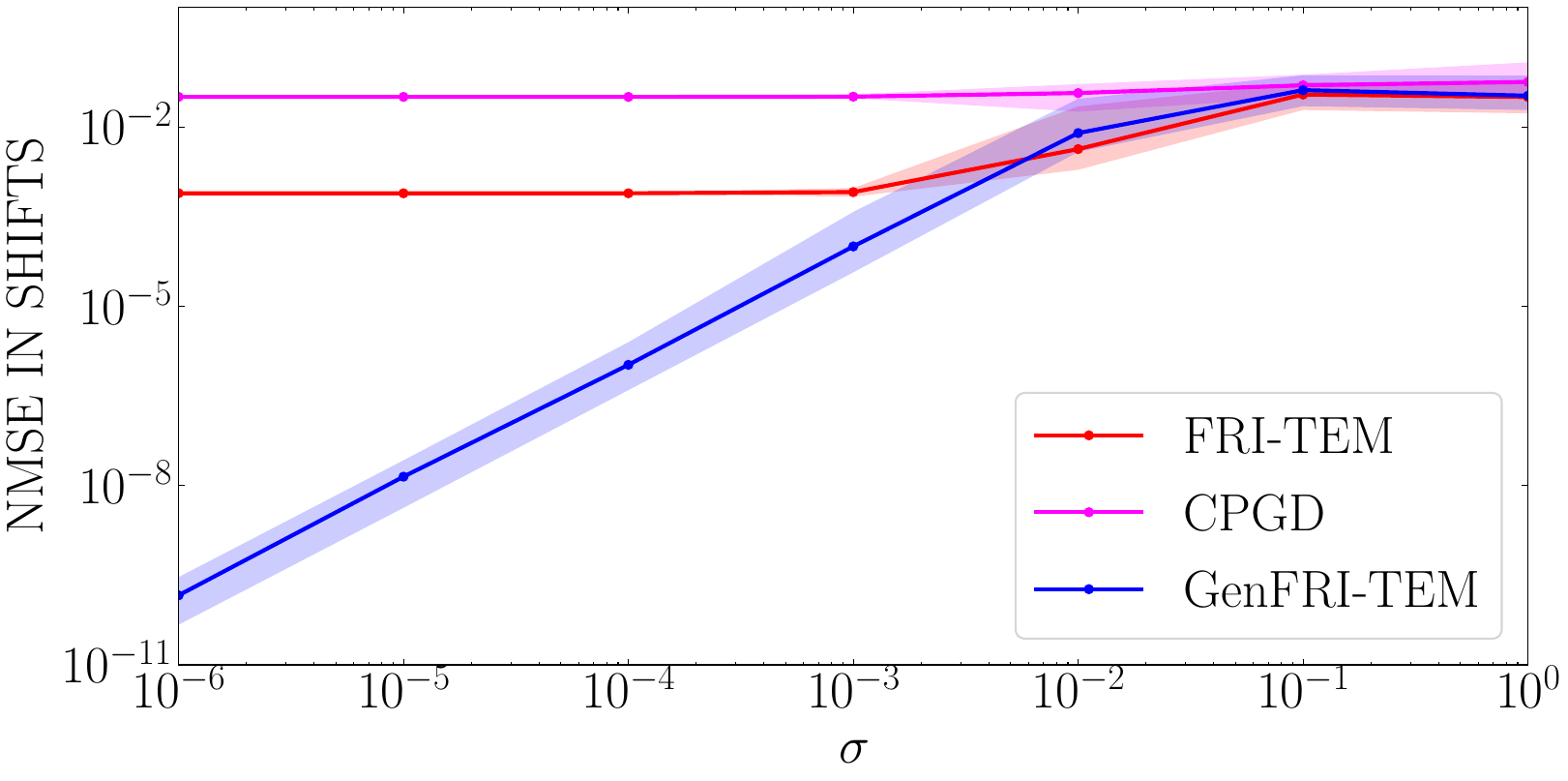}}
\subfigure[]{\label{fig:fitem_amp_nmse}\includegraphics[width=3.5in]{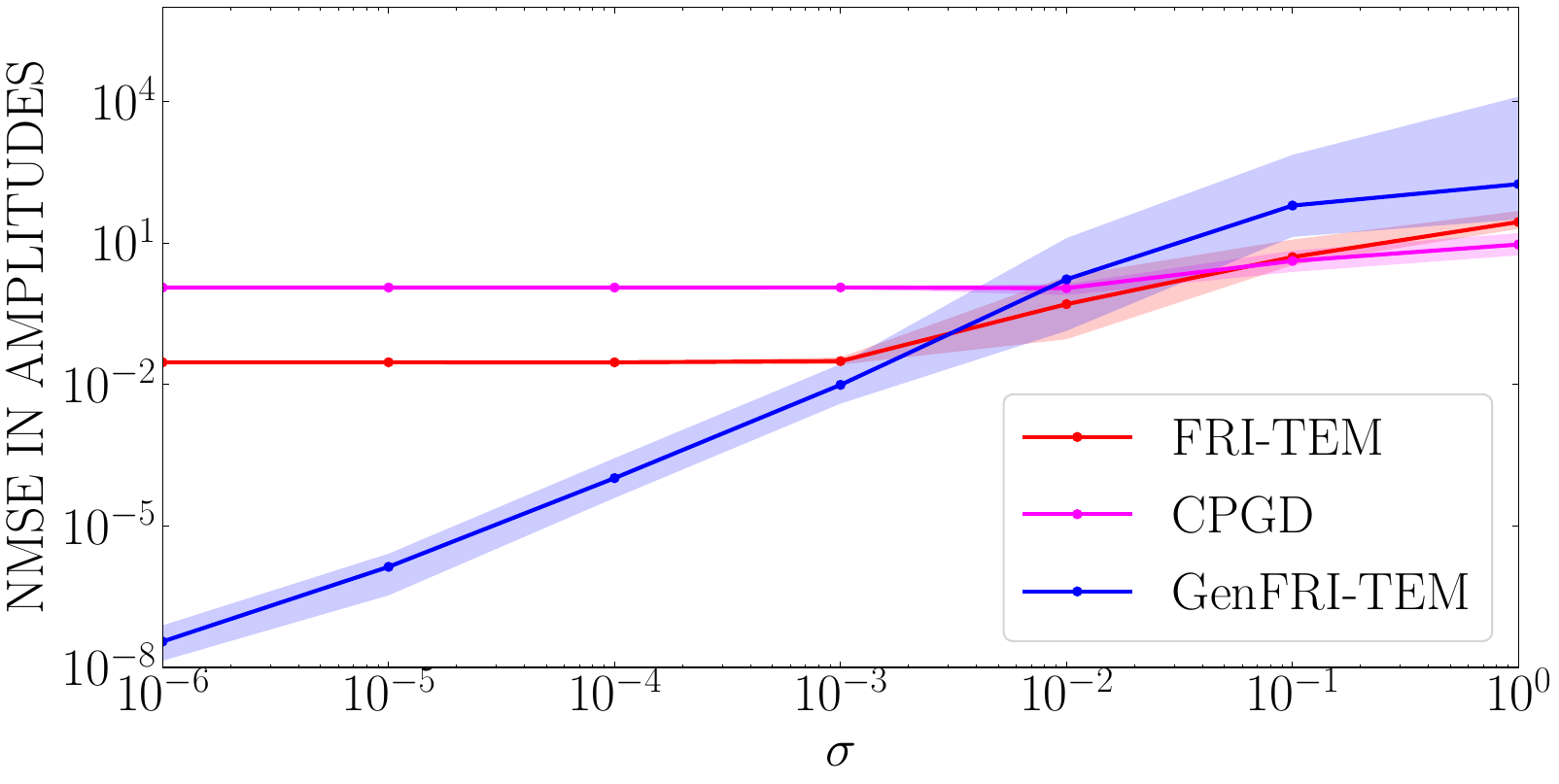}}
\caption{Normalized mean-square error in estimating the amplitude and shift parameters as a function of the noise level $\sigma$ from IF-TEM measurements. The solid line shows the median error and the shaded region captures the inter-quartile region of the estimate. The results are over $100$ realizations of the jitter. The axes are in log scale.}
\label{fig:iftem_errors_noisy}
\end{figure}
\begin{figure}[t]
\centering
\subfigure[$\sigma=10^{-6}$]{\label{fig:}\includegraphics[width=1.72in]{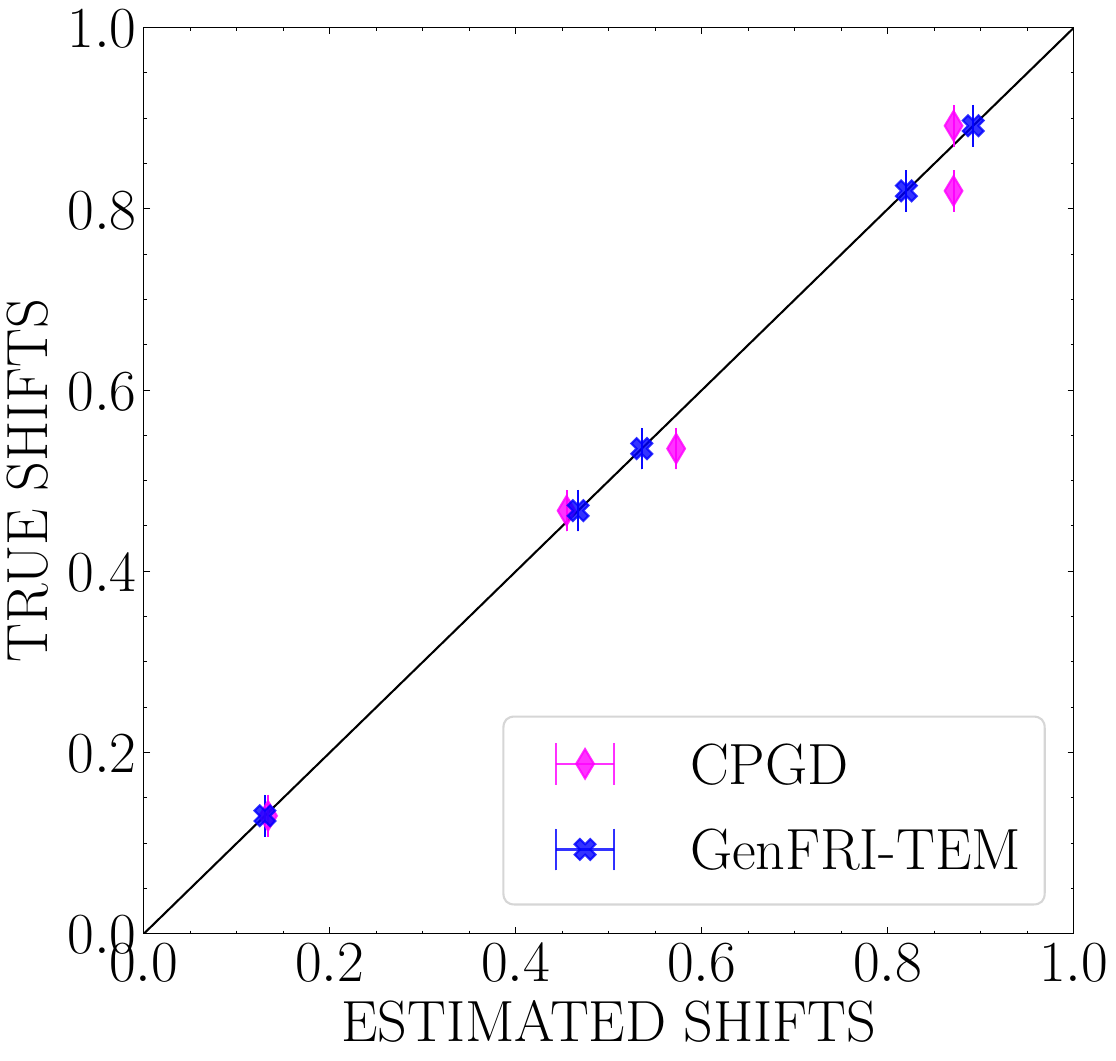}}
\subfigure[$\sigma=10^{-5}$]{\label{fig:}\includegraphics[width=1.72in]{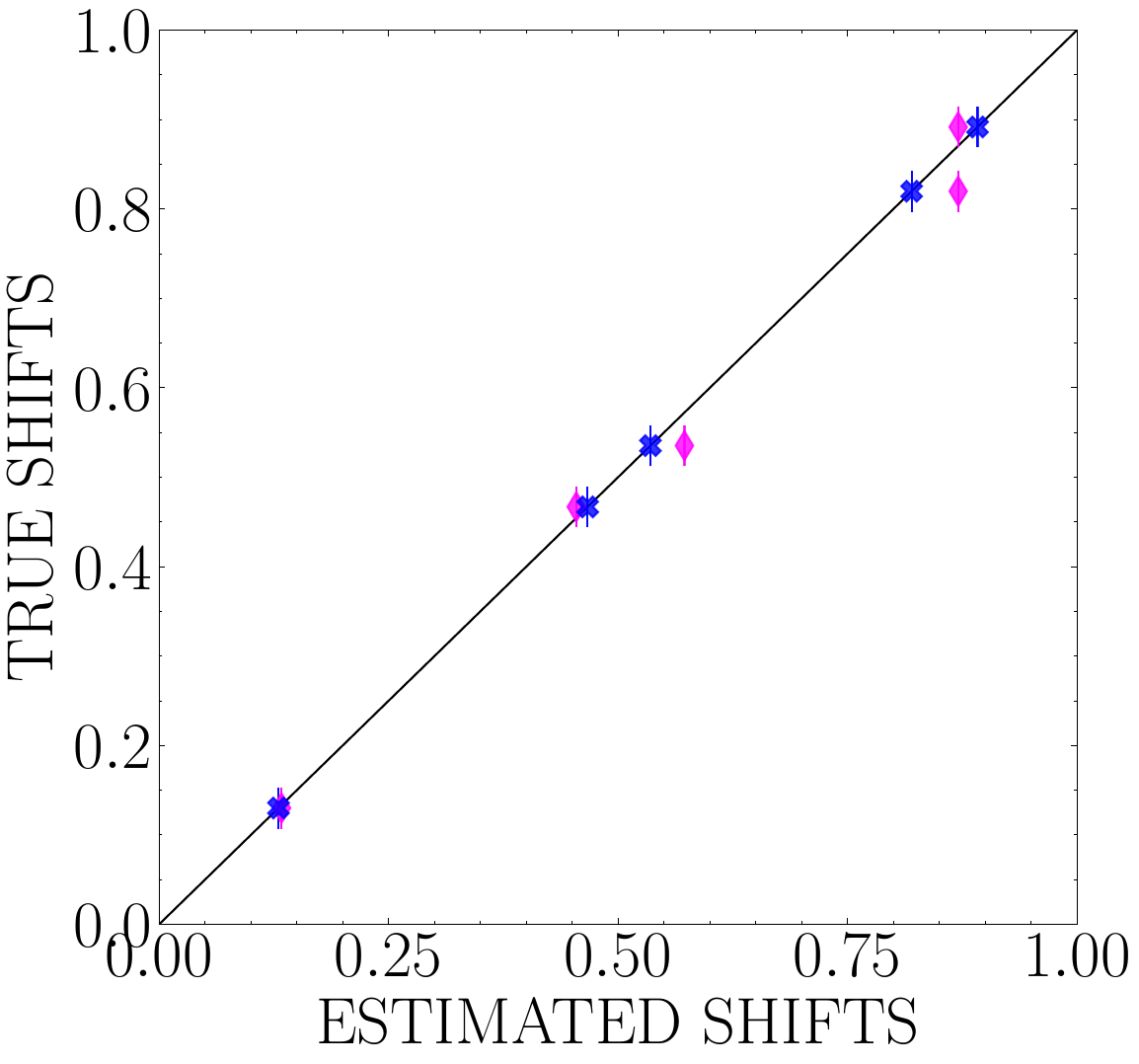}}
\subfigure[$\sigma=10^{-2}$]{\label{fig:}\includegraphics[width=1.72in]{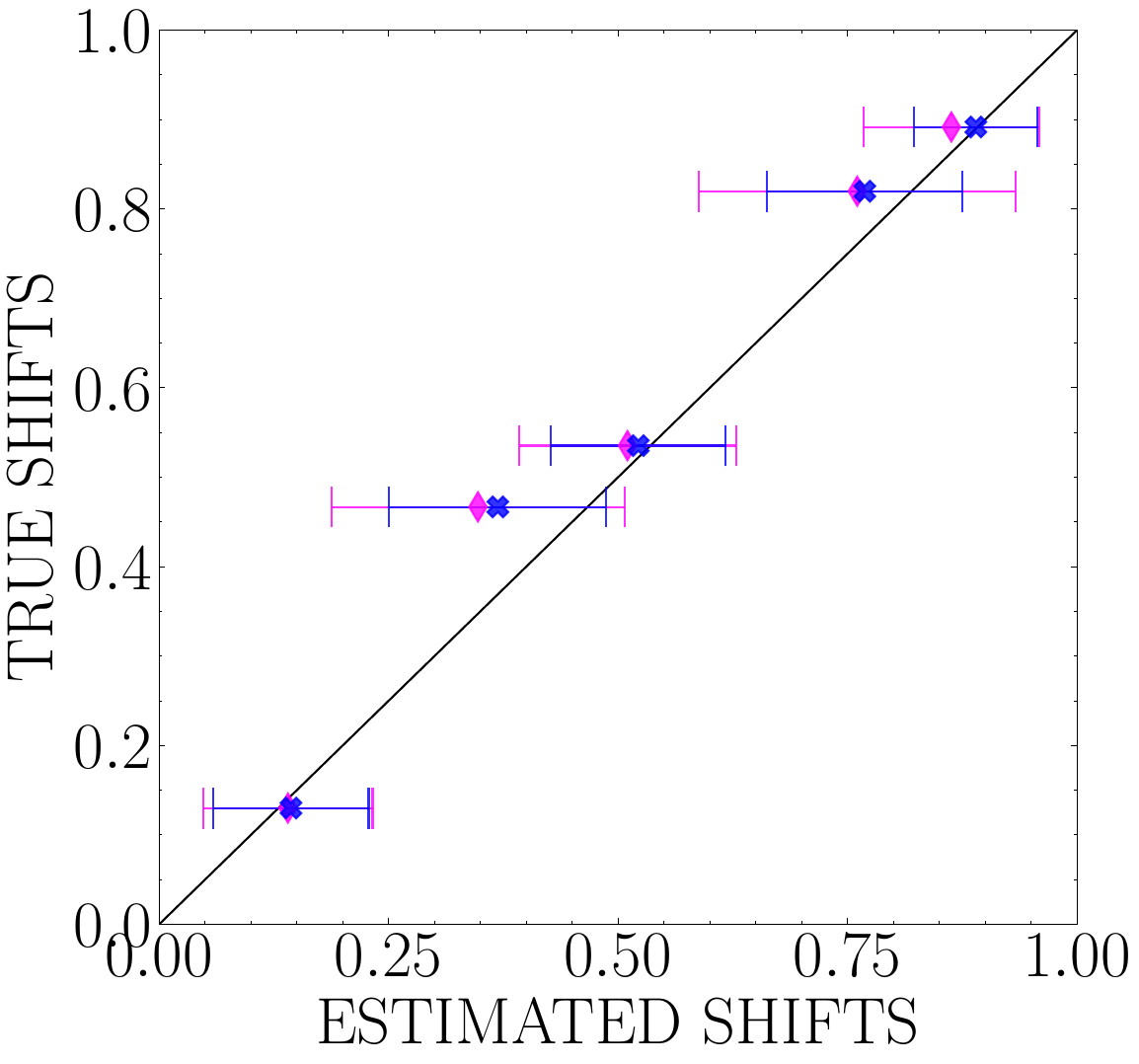}}
\subfigure[$\sigma=1$]{\label{fig:}\includegraphics[width=1.72in]{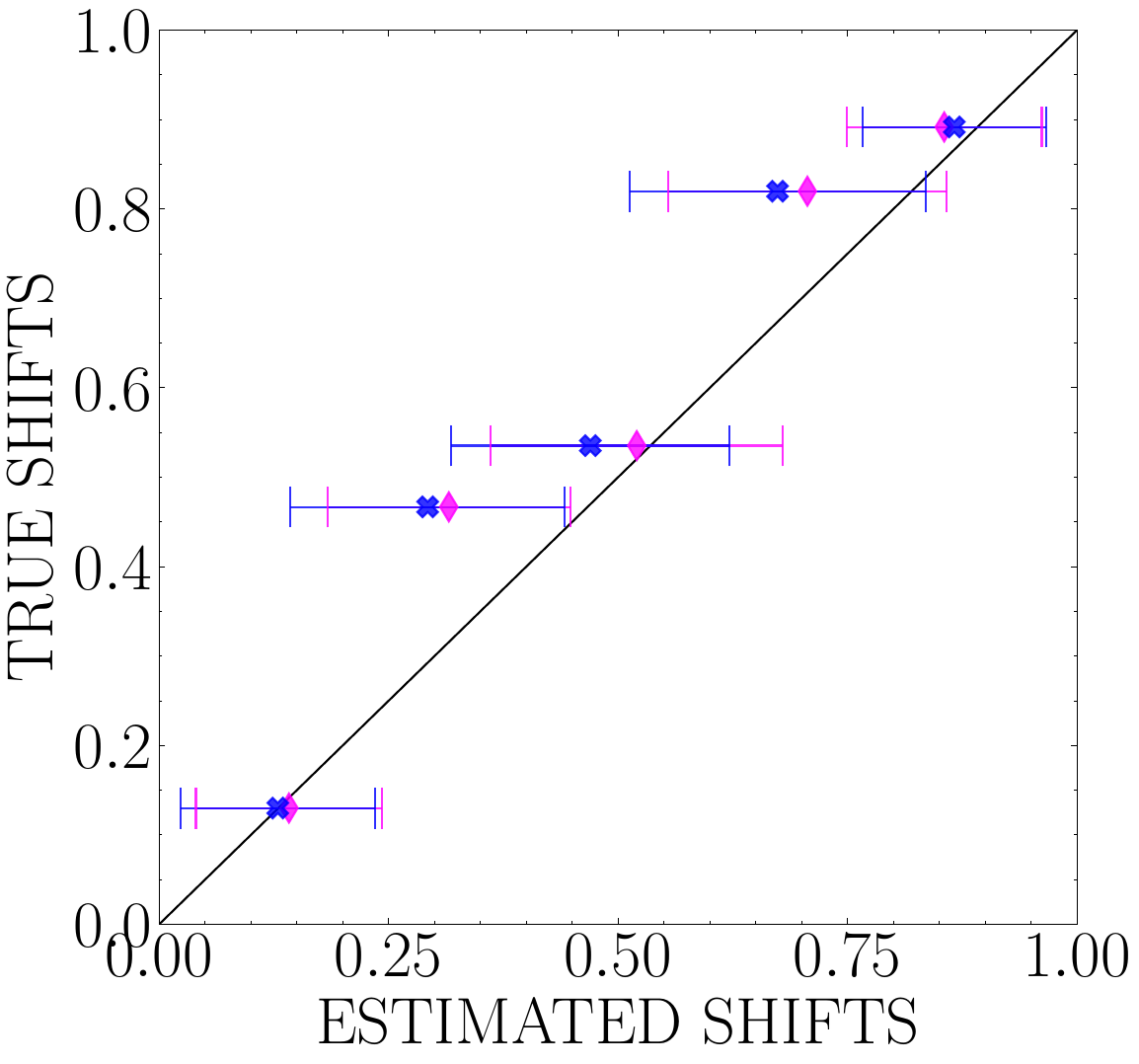}}
\caption{Comparison of estimated shifts from C-TEM measurements with the true shifts, for varying values of jitter level $\sigma$. Each marker corresponds to the average estimated shift with the error bar denoting one standard deviation. The closer the marker is to the $45^{\circ}$ straight line, the better is the recovery.}
\label{fig:ctem_position_noisy}
\end{figure}
\begin{figure}[t]
\centering
\subfigure[$\sigma=10^{-6}$]{\label{fig:}\includegraphics[width=1.72in]{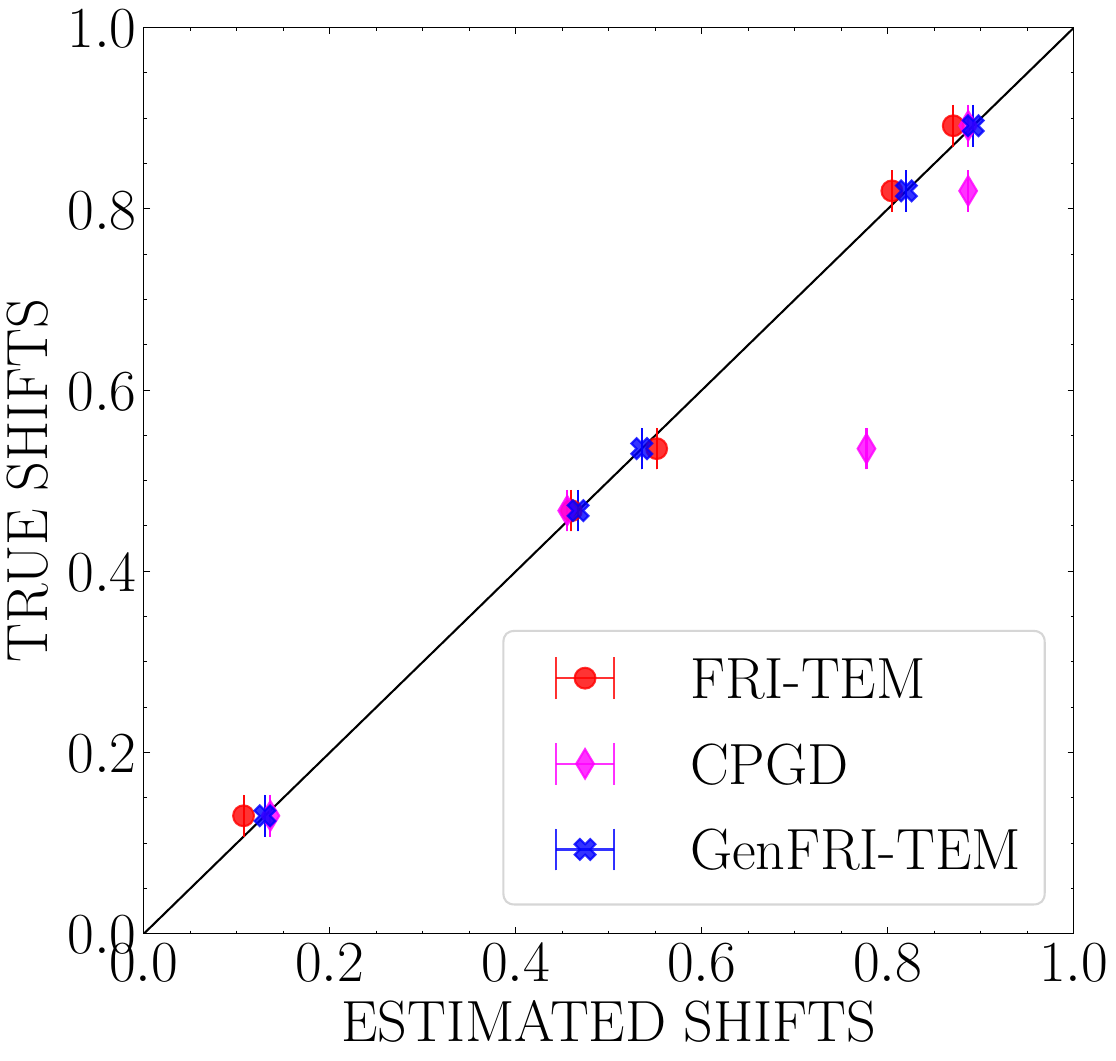}}
\subfigure[$\sigma=10^{-5}$]{\label{fig:}\includegraphics[width=1.72in]{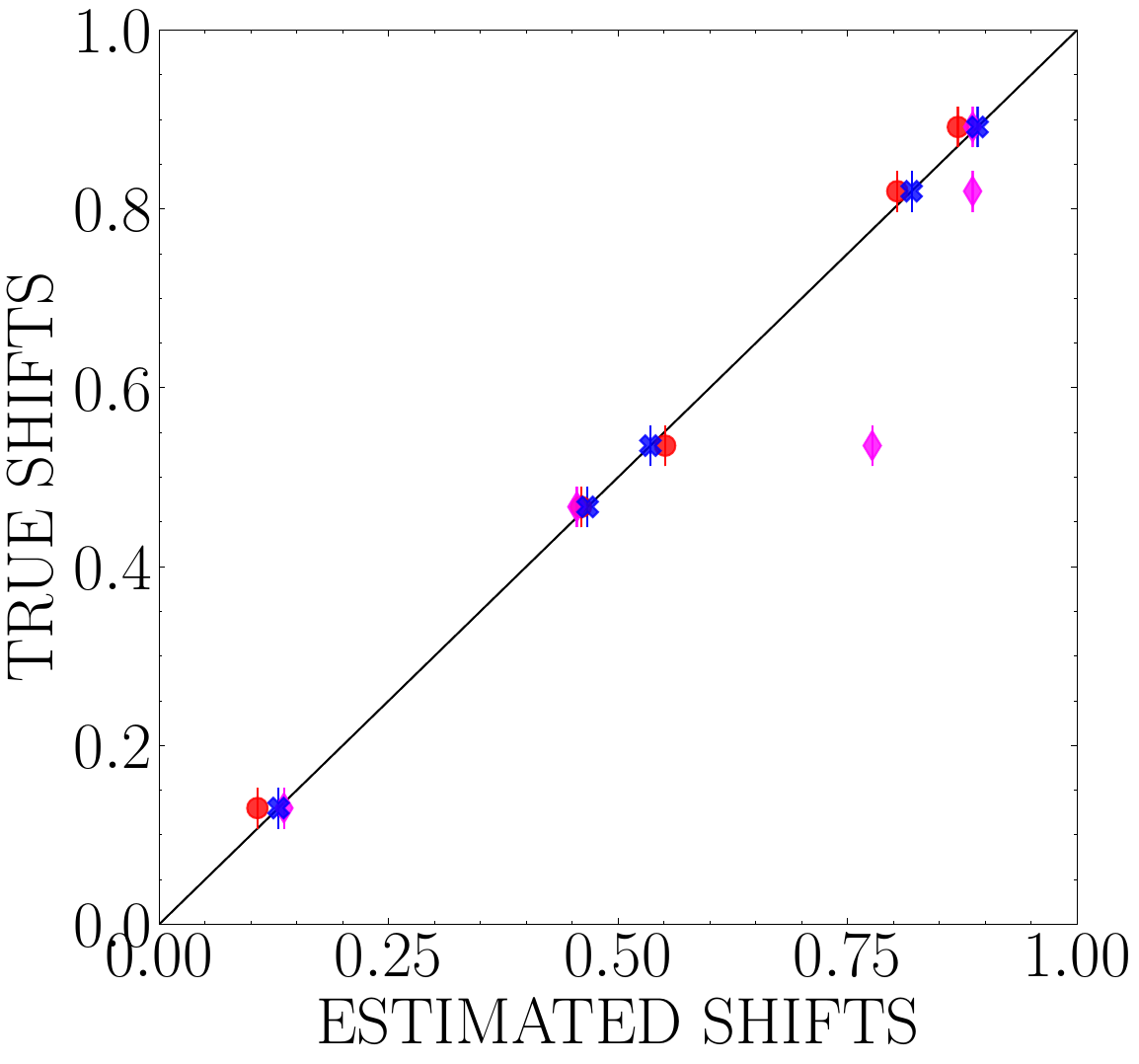}}
\subfigure[$\sigma=10^{-2}$]{\label{fig:}\includegraphics[width=1.72in]{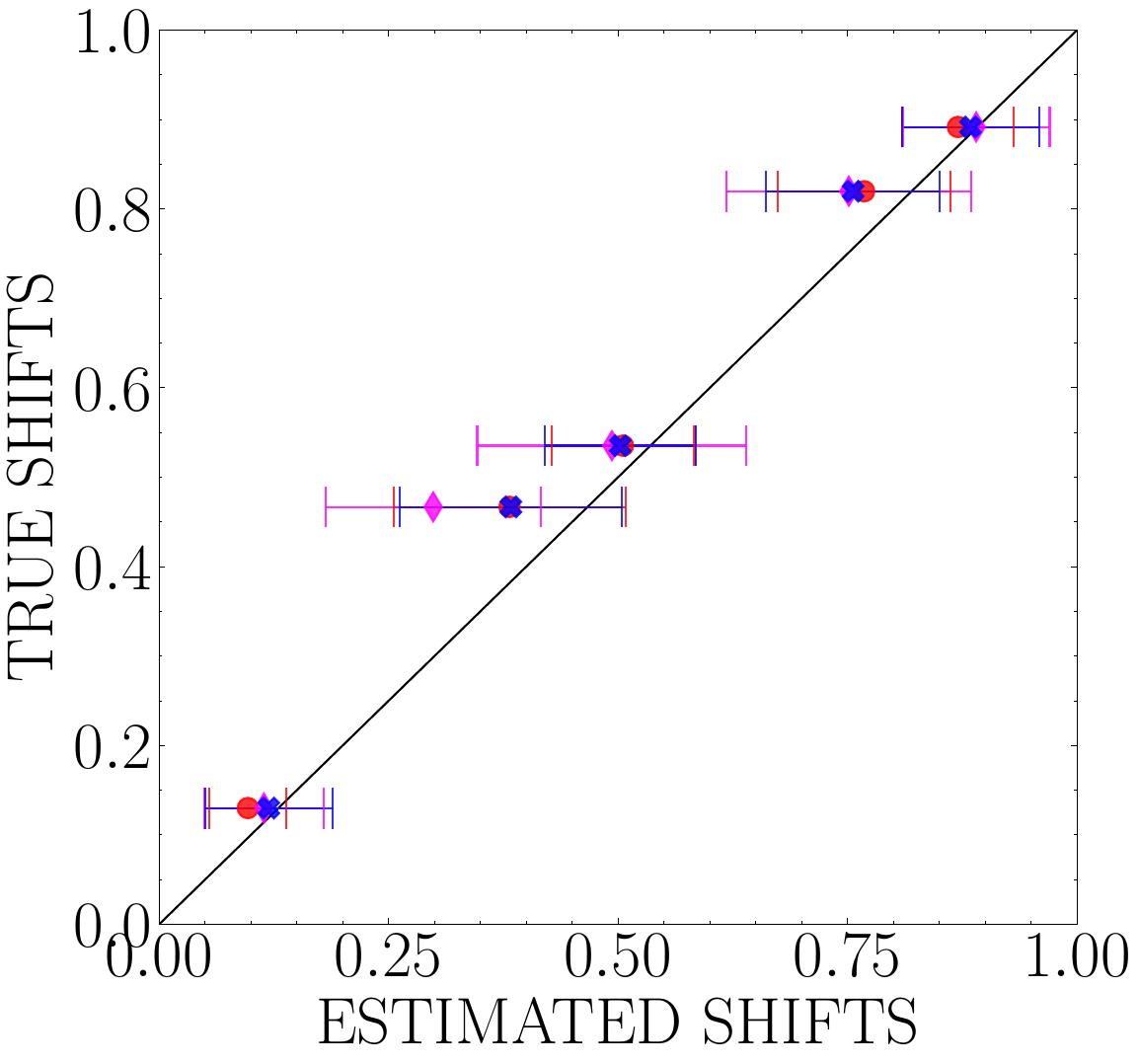}}
\subfigure[$\sigma=1$]{\label{fig:}\includegraphics[width=1.72in]{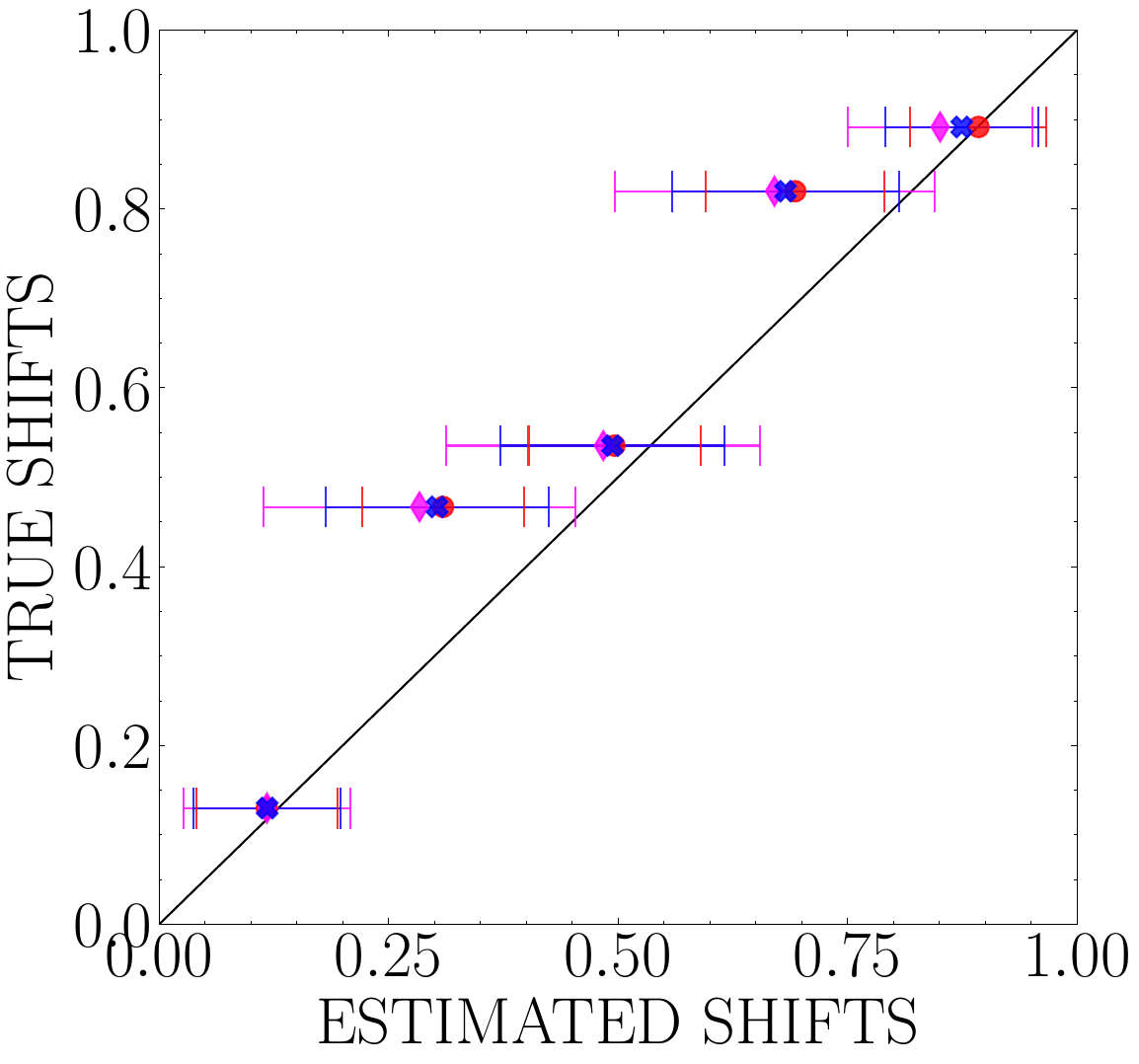}}
\caption{Comparison of estimated shifts from IF-TEM measurements with the true shifts, for varying values of jitter level $\sigma$. Each marker corresponds to the average estimated shift with the error bar denoting one standard deviation. The closer the marker is to the $45^{\circ}$ straight line, the better is the recovery.}
\label{fig:iftem_position_noisy}
\end{figure}
\subsection{Reconstruction from Noisy Measurements}
\label{subsec:noisy_experiments}
We next consider the effect of measurement noise in the form of jitter in the trigger times. The jitter is modelled with $\sigma=10^{-3}$ (cf. Eq.~\eqref{eq:noiseModel}). Figures~\ref{fig:ctem_noise_rec}~and~\ref{fig:iftem_noise_rec} show the signal $x(t)$ and the median of the reconstruction $\check{x}(t)$ from noisy C-TEM and IF-TEM measurements, respectively, over $100$ realizations of jitter. Let $\{(\check{c}_{k},\check{\tau}_{k})\}_{k=0}^{K-1}$ denote the estimated amplitudes and shifts. The normalized mean-squared error (NMSE) in estimating the parameters is given by
\begin{equation}
\begin{split}
	\text{NMSE}_{\boldsymbol{c}} = \frac{\sum_{k=0}^{K-1} (c_{k} - \check{c}_{k})^{2}}{\sum_{k=0}^{K-1} c_{k}^{2}},
	\text{NMSE}_{\boldsymbol{\tau}} = \frac{\sum_{k=0}^{K-1} (\tau_{k} - \check{\tau}_{k})^{2}}{\sum_{k=0}^{K-1} \tau_{k}^{2}}.
\end{split}
\label{eq:mseShifts}
\end{equation}
In both the cases, the $\text{NMSE}_{\boldsymbol{\tau}}$ is of the order of $10^{-5}$, and $\text{NMSE}_{\boldsymbol{c}}$ is of the order of $10^{-3}$, indicating accurate reconstruction. \\
\indent Next, we perform Monte Carlo simulations to assess the performance for different levels of jitter: $\sigma = 10^{-6}, 10^{-5}, 10^{-4}, 10^{-3}, 10^{-2}, 10^{-1}, 10^{0}$. Higher the $\sigma$, more is the jitter. We consider reconstruction using GenFRI-TEM (Algorithm~\ref{algo:genFRI}), FRI-TEM \cite{naaman2021fritem} and Cadzow Plug-and-Play Gradient Descent (CPGD) \cite{simeoni2020cpgd,simeoni2020pyoneer} techniques. The NMSE in shifts and amplitudes (Eq.~\eqref{eq:mseShifts}) is averaged over $100$ realizations of the jitter for both C-TEM and IF-TEM. Figures~\ref{fig:ctem_errors_noisy}~and~\ref{fig:iftem_errors_noisy} shows the NMSE as a function of $\sigma$ for C-TEM and IF-TEM, respectively. For small $\sigma$, GenFRI-TEM provides the least NMSE compared with the benchmark methods. For large $\sigma$, all techniques have comparable NMSEs.\\
\indent The superior performance of GenFRI-TEM for low $\sigma$ is attributed to the fact that it jointly solves for the annihilating filter and the Fourier coefficients subject to the annihilation constraint $(\Gamma_K\hat{\bx})\bh = \zerovec$, whereas CPGD and FRI-TEM solve for them sequentially. Consequently, errors in the estimation of the Fourier coefficients will lead to a full-rank $\Gamma_K\hat{\bx}$ and errors in the estimation of the shifts. \\
\indent Figures~\ref{fig:ctem_position_noisy}~and~\ref{fig:iftem_position_noisy} show the average of the estimated shifts compared with the true shifts for C-TEM and IF-TEM, respectively, for various $\sigma$. The recovery of shifts is accurate if the mean is close to the $45^\circ$ straight line, and the standard deviation (indicated by an error bar) is small. Deviation of the average from the straight line indicates a bias in the recovery. \\
\indent For lower $\sigma$, CPGD and GenFRI-TEM have less bias, while FRI-TEM shows outliers. For large $\sigma$, the estimation bias is high in general for all techniques, but the bias is the least with GenFRI-TEM. The standard deviation of the estimated shifts using GenFRI-TEM can also be observed to be smaller than the benchmark methods. This is consistent with the results in Figure~\ref{fig:ctem_errors_noisy}~and~\ref{fig:iftem_errors_noisy}. This evidence shows that GenFRI-TEM has least NMSE and least bias.\\
\indent The results pertaining to time-encoding of FRI signals generated by a cubic B-spline kernel are reported in the Supplementary Material.

%% -----------------------------------------------------------
%					CONCLUSIONS
%% -----------------------------------------------------------

\section{Conclusions}
\label{sec:conclusions}
We considered time-encoding of FRI signals, in particular, using the crossing-time-encoding and the integrate-and-fire time-encoding machines. We analyzed time-encoding in the Fourier domain, and showed the time-encoded measurements of the FRI signal, obtained via a suitable sampling kernel, can be used to compute the Fourier coefficients by solving a linear system of equations. The framework extends to the multichannel setting, where the sampling requirement is reduced in each channel. Using the Fourier coefficients, standard FRI signal reconstruction techniques become applicable. We presented sufficient conditions for perfect reconstruction using single- and multichannel C-TEM and IF-TEM. We also considered the effect of measurement noise as jitter in the trigger times. We proposed an optimization strategy for signal reconstruction, which is similar to the Generalized FRI technique (GenFRI-TEM) to jointly solve for the Fourier coefficients and the annihilating filter. Simulation results showed that GenFRI-TEM provides the least mean-squared error and the lowest bias in signal reconstruction compared with the benchmark methods.

%% -----------------------------------------------------------
%					APPENDIX
%% -----------------------------------------------------------

\section*{Acknowledgments}
This work was supported by Pratiksha Trust and Indian Institute of Science under the Institute of Eminence project.

\appendices
\section{Proof of Lemma~\ref{lem:ctemSamplingDensity}}
\label{appendix:proofLemmactemSamplingDensity}
\begin{proof}
Let $y: [0, T[ \rightarrow \rr$ be a continuous and differentiable signal encoded using a C-TEM with reference $r(t) = A_{r}\cos(2\pi f_{r}t)$, where $A_{r}\geq \Vert y \Vert_{\infty}$. Consider the partitioning of $[0, T]$ into smaller disjoint intervals $\mathcal{U}_{m} = \left[m\frac{1}{2f_r}, (m+1)\frac{1}{2f_r}\right[$, $m = 1,2,\cdots$. For every $m$, we have, $\left( r\left(m\frac{1}{2f_r}\right) - y\left(m\frac{1}{2f_r}\right) \right)  \cdot \left( r\left((m+1)\frac{1}{2f_r}\right) - y\left((m+1)\frac{1}{2f_r}\right) \right) < 0$, i.e., the sign of the difference signal $(r-y)$ changes once in the interval $\mathcal{U}_{m}$. Since $r$ is continuous and differentiable, so is $r-y$, and hence, using Bolzano's intermediate value theorem (cf. Chapter 9, \cite{tao2009analysis}), $\exists \; t_n \in \mathcal{U}_{m}$ such that $r(t_n) - y(t_n) = 0$, i.e., in the interval of half a period of the reference $r$, the signal $(y-r)$ crosses zero at least once at $t_n$. Therefore, $\displaystyle \sup_{n\in\zz} \vert t_{n+1} - t_{n} \vert < \frac{1}{f_r}$.
\end{proof}

%% -----------------------------------------------------------

\section{Proof of Lemma~\ref{lem:ctemMatrix}}
\label{appendix:proofLemmactemMatrix}
\begin{proof}
Consider the case where $L=N$, i.e., the matrix $\bG_{\mathrm{CT}}$ is a square matrix. We will show that the matrix has a nonzero determinant, and $\rank(\bG_{\mathrm{CT}}) = N$. For $L > N$, addition of rows to the $N\times N$ matrix cannot decrease the rank. Using the properties of determinants:
\begin{equation}
\det(\bG_{\mathrm{CT}}) = \det(\bF)\cdot \prod_{n=1}^L e^{-\jj M \omega_0 t_n},
\label{eq:prodGct}
\end{equation}
where %the matrix $\bF \in \cc^{L\times N}$ is given by
\begin{equation*}
\bF =
\begin{bmatrix}
1  & e^{\mathrm{j}\omega_0 t_{1}} &\hspace{-0.2cm}\cdots &\hspace{-0.1cm}e^{\mathrm{j}2M\omega_0 t_{1}}\\
1 & e^{\mathrm{j}\omega_0 t_{2}} &\hspace{-0.2cm}\cdots &\hspace{-0.1cm}e^{\mathrm{j}2M\omega_0 t_{2}}\\
\vdots & \vdots & & \vdots\\
1 & e^{\mathrm{j}\omega_0 t_{L}} &\hspace{-0.2cm}\cdots &\hspace{-0.1cm}e^{\mathrm{j}2M\omega_0 t_{L}}  \/
\end{bmatrix},
\end{equation*}
which is a Vandermonde matrix. By Definition~\ref{def:tem}, the set $\{t_n\}_{n=1}^L$ has increasing and distinct entries. Further, since $0 \leq t_1 < t_2 < \cdots < t_L \leq T$, the rows of $\bF$ are distinct. Using the properties of Vandermonde matrices \cite{horn2012matrix}, $\det(\bF) \neq 0$ whenever $L \geq N$ and each term in the product in Eq.~\eqref{eq:prodGct} is nonzero. Hence, $\det(\bG_{\mathrm{CT}}) \neq 0$ whenever $L \geq N$.
\end{proof}

%% -----------------------------------------------------------

\section{Proof of Proposition \ref{prop:ctemGuarantees}}
\label{appendix:proofPropCtemGuarantee}
\begin{proof}
Let $x(t)$ be a $T$-periodic FRI signal time-encoded using a sampling kernel $g(t)$ and a C-TEM with reference $r(t) = A_{r}\cos(2\pi f_{r}t)$. The signal is characterized by the $2K$ parameters $\{(c_{k}, \tau_{k})\}_{k=0}^{K-1}$, and from Section~\ref{subsec:pronysmethod}, we know that the parameters can be recovered using $2K+1$ Fourier coefficients. Let the sampling kernel $g(t)$ satisfy Eq.~\eqref{eq:samplingKernel}. Then, from Eq.~\eqref{eq:CTEMmeasurements}, setting $M=K$, we have $2K+1$ contiguous Fourier coefficients satisfying a linear system of equations. Using Lemma~\ref{lem:ctemMatrix}, the linear system has a unique left-inverse when $L\geq 2K+1$. Therefore, the C-TEM must sample such that $L\geq 2K+1$. \\
\indent Let the amplitude of the reference of the C-TEM satisfy $A_{r} \geq \Vert x*g \Vert_{\infty}$. Lemma~\ref{lem:ctemSamplingDensity} ensures that there is at least one measurement in every interval $\left[m\frac{1}{2f_{r}}, (m+1)\frac{1}{2f_{r}} \right[, m=1,2,\cdots$. Therefore, obtaining $2K+1$ measurements in the interval $[0,T[$ requires that $\displaystyle (2K+1)\frac{1}{f_{r}} < T$. Further, with $L\geq (2K+1)$, using Lemma~\ref{lem:ctemMatrix}, $\bG_{\mathrm{CT}}$ has a left-inverse. The $(2K+1)$ Fourier coefficients are sufficient for estimation of the unknown support and amplitudes using Prony's method (Section~\ref{subsec:pronysmethod}).
% \begin{remark}
% The last part of this proof and Lemma~\ref{lem:ctemSamplingDensity} do not require the FRI structure on the input signal, other than the fact that $2K+1$ samples are required.
% \end{remark}
\end{proof}

%% -----------------------------------------------------------

\section{Proof of Lemma~\ref{lem:iftemMatrix}}
\label{appendix:proofLemmaiftemMatrix}
\begin{proof}
Consider the matrix $\bG_{\mathrm{IF}}$ as defined in Eq.~\eqref{eq:iftemMatrix}. From  Eq.~\eqref{eq:iftemMeasurements}, the matrix can be viewed as the result of finite-difference operation on a matrix of the type $\bG_{\mathrm{CT}}$. Specifically, we can write $\bG_{\mathrm{IF}}$ as a product of the finite-difference matrix:
\begin{equation}
	\mathbf{D} = \begin{bmatrix}
		1 & -1 & \cdots & 0 & 0 \\
		0 & 1 & \cdots & 0 & 0 \\
		\vdots & \vdots &\, \ddots & \vdots & \vdots \\
		0 & 0 & \cdots & 1 & -1
	\end{bmatrix} \in \cc^{(L-1)\times L},
\end{equation}
and the matrix $\tilde{\bG}_{\mathrm{CT}}\in\cc^{L\times N}$ given by
\begin{equation}
	\tilde{\bG}_{\mathrm{CT}} = \begin{bmatrix}
		e^{-\jj M\omega_{0}t_{1}} & \cdots & t_{1} & \cdots & e^{\jj M\omega_{0}t_{1}} \\
		\vdots &\, \ddots & \vdots &\, \ddots & \vdots \\
		e^{-\jj M\omega_{0}t_{L}} & \cdots & t_{L} & \cdots & e^{\jj M\omega_{0}t_{L}}
	\end{bmatrix}.
\end{equation}
We are interested in solving the equation $\bG_{\mathrm{IF}}\bu = \mathbf{D}\tilde{\bG}_{\mathrm{CT}}\bu = \zerovec$, and would like to show that the set of all solutions $\bu$ is a singleton containing only the all-zero vector. \\
\indent The dimension of the null space of the finite-difference matrix is $1$, and has vectors of the form $\alpha \mathbbm{1}$, where $\alpha\in\cc$ and $\mathbbm{1}\in\cc^{L}$ is the vector of all ones. Therefore, the problem reduces to finding the set of vectors $\bu = [u_{-M} \; \cdots \; u_{M}]^\TT$ that satisfy $\tilde{\bG}_{\mathrm{CT}}\bu = \alpha\mathbbm{1}$. Any entry in the equation has the form:
\begin{equation}
	\sum_{\substack{m=-M \\ m \neq 0}}^{M} u_{m}e^{\jj m\omega_{0}t_{\ell}} + u_{0}t_{\ell} = \alpha.
\label{eq:lineCondition}
\end{equation}
The set of vectors $\bu$ that satisfy Eq.~\eqref{eq:lineCondition} for $\ell = 1,2,\cdots,L$ is empty for $L>N$. This is because the solutions, with rearrangement, can be written as the solutions of the coefficients of an $M^{\text{th}}$ degree trigonometric polynomial that intersects a straight line exactly at $t_{1},t_{2},\cdots,t_{L}$. This problem has no solution for $L>N$, as the $M^{\text{th}}$ degree polynomial can have only exactly $N=2M+1$ such points. Naaman {\it et al.} have a related result in their recent preprint \cite{naaman2021fritem}.
\end{proof}

%% -----------------------------------------------------------

\section{Proof of Proposition \ref{prop:iftemGuarantees}}
\label{appendix:proofPropIFtemGuarantee}
\begin{proof}
Let $x(t)$ be a $T$-periodic FRI signal that is time-encoded using a sampling kernel $g(t)$ and an IF-TEM with parameters $\{b,\kappa,\gamma\}$ (cf. Figure~\ref{fig:schematicKernelBasedSampling}). The filtered signal $y(t) = (x*g)(t)$ is the input to the IF-TEM. Let $t_{1} \in [0,T[$ be the first trigger time. Obtaining $(L-1)$ trigger times in $[0,T[$ requires (using Corollary~\ref{cor:iftemSamplingDensity}) that
\begin{equation}
	t_{1} + (L-1) \frac{\kappa\gamma}{b-\Vert y \Vert_{\infty}} < T.
\label{eq:iftemSamplingCondition}
\end{equation}
Without loss of generality, assume zero-initial condition for the integrator: $v(0) = 0$. The first trigger time is bounded as follows:
\begin{equation*}
\begin{split}
	\gamma = v(t_{1}) &= \frac{1}{\kappa}\int_{0}^{t_{1}} \min_{\nu} (b+y(\nu))\, \dd \nu, \\
	&\geq \frac{1}{\kappa}\int_{0}^{t_{1}} \left(b-\Vert y \Vert_{\infty}\right)\, \dd\nu = \frac{1}{\kappa} \left(b-\Vert y \Vert_{\infty}\right) t_{1}, \\
	& \implies t_{1} \leq \frac{\kappa\gamma}{b-\Vert y \Vert_{\infty}}.
\end{split}
\end{equation*}
The above inequality taken together with Eq.~\eqref{eq:iftemSamplingCondition} gives the required condition in the Proposition.
% \begin{equation}
% 	\frac{\kappa\gamma}{b-\Vert x*g \Vert_{\infty}} < \frac{T}{L},
% \label{eq:iftemSamplingCondition2}
% \end{equation}
% i.e., to obtain $L\geq (2K+1)$ time instants in $[0,T[$ Eq.~\eqref{eq:iftemSamplingCondition2} must hold.
Further, with $L\geq (2K+1)$, using Lemma~\ref{lem:iftemMatrix}, $\bG_{\mathrm{IF}}$ has a left-inverse. $(2K+1)$ Fourier coefficients are sufficient for estimation of the unknown support and amplitudes using Prony's method (Section~\ref{subsec:pronysmethod}).
\end{proof}

\bibliographystyle{IEEEbib}
\bibliography{references.bib}

\includepdf[pages=-]{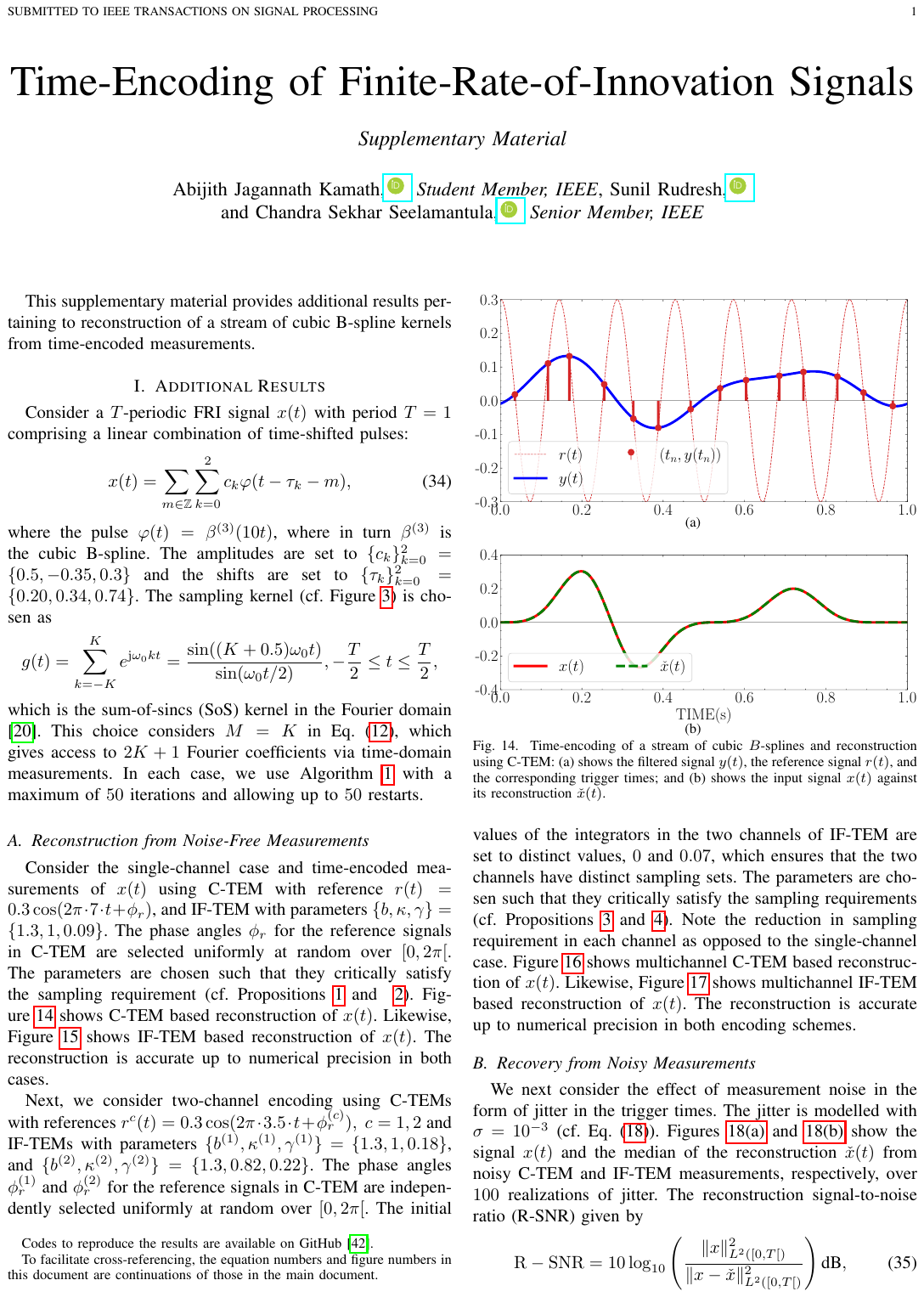}

\end{document}